\documentclass[11pt]{article}

\usepackage{amsmath}
\usepackage{graphicx}
\usepackage{indentfirst}
\usepackage{amssymb}
\usepackage{cite}
\usepackage{color}
\usepackage{subfigure}
\usepackage{varwidth}

\begin{document}

\begin{center}
\large {\bf Holographic Einstein rings of a Gauss-Bonnet AdS black hole}
\end{center}

\begin{center}
Xiao-Xiong Zeng $^a$$^{,b}$
 $\footnote{E-mail: \texttt{xxzengphysics@163.com}}$,
  Ke-Jian He $^{c}$
 $\footnote{E-mail: \texttt{kjhe94@163.com}}$,
  Jin Pu $^{d}$
 $\footnote{E-mail: Corresponding author, \texttt{pujin@cwnu.edu.cn}}$
 and Guo-Ping Li $^{d}$
 $\footnote{E-mail: Corresponding author, \texttt{gpliphys@yeah.net}}$
\end{center}

\begin{center}
\textit{a. State Key Laboratory of Mountain Bridge and Tunnel Engineering, Chongqing Jiaotong
University, Chongqing 400074, People's Republic of China\\
b. Department of Mechanics, Chongqing Jiaotong University, Chongqing 400074, People's Republic of China\\
c. College Of Physics, Chongqing University, Chongqing 401331, People's Republic of China \\
d. School of Physics and Astronomy, China West Normal University, Nanchong 637000, People's Republic of China
}
\end{center}
\pdfoutput=1

\noindent
{\bf Abstract:}
Based on the AdS/CFT correspondence, we studied the holographic Einstein images of a Gauss-Bonnet AdS black hole in the bulk from a given response function on AdS boundary. For the absolute amplitude of total response function, it shows that there always exists the interference pattern as the scalar wave passed through the black hole. And, the absolute amplitude depends closely on the properties of Gaussian source and spacetime geometry. More importantly, we also find that the holographic images always appears as a ring with the concentric stripe surrounded when the observer located at the north pole. At other positions, this ring will change into a luminosity-deformed ring, or two light points. In addition, the effects of Gauss-Bonnet parameter $\alpha$, wave source and optical system on the holographic images have been carefully addressed throughout of paper. Finally, we conclude that the holographic images in different types of black holes with some different features may shed some deep insight on the existence of a gravity dual for a given material.

\thispagestyle{empty}
\newpage
\setcounter{page}{1}

\section{Introduction}
\label{sec:intro}
The AdS/CFT correspondence as a concrete realization of holographic principle explicitly identifies that the theory of quantum gravity in Anti-de-Sitter space-time(AdS) is equivalent to a dual conformal quantum field theory (CFT)\cite{maldacena1999AdSCFT}. Among them, the famous example is the pair between the type $IIB$ string theory on $AdS_5 \times S^5$ and the maximally supersymmetric gauge theory in four dimensions, i.e., $\mathcal{N}=4$ super-Yang-Mills (SYM)\cite{aharony2000large,natsuume2015ads}. Since then, the holographic property of gravity has been widely accepted and applied to many studies in various fields of physics.
At present, this correspondence is regarded as a useful tool to deal with some problems faced by strong coupling systems. For instance, some QCD-like models are constructed through the application of AdS/CFT duality in low energy quantum chromodynamics (QCD). And, many properties of the strong coupling region of QCD are studied using these holographic QCD models, such as constrained phase transition, chiral phase transition, and QCD vacuum \cite{erlich2005QCD}. In addition, the application of AdS/CFT correspondence in condensed matter physics has also attracted extensive attention\cite{hartnoll2009CMT}, especially in superfluidity, superconductivity, non Fermi liquids and Fermi liquids, which shed some new lights for understanding high-temperature superconducting physics\cite{gubser2008breaking,hartnoll2008HS,hartnoll2008building,herzog2009holographic}. Also, in spirit of AdS/CFT, some other holographic correspondences have also been further studied, such as dS/CFT correspondence and Kerr/CFT correspondence \cite{strominger2001ds, bredberg2011lectures}.
So far, the holographic principle attracted much attention of physicists, and it has become an effective tool to study various physical topics in the background of modified gravity\cite{Huang:2004ai,Li:2009zs,Sheykhi:2009zv,Micheletti:2009jy,Setare:2010wt,Huang:2010zzt,Lu:2009iv,Bai:2014poa,Aprile:2012sr,Cai:2017ihd,Kusuki:2019zsp,Akers:2019nfi,Bhattacharya:2021jrn,Karndumri:2022rlf}.

Black hole is one of the most interesting celestial body, which predicted by general relativity (GR). And recently, it has been proved to be existed in the universe. With the development of technology, the gravitational wave detection results provided by the Laser Interferometer Gravitational Wave Observatory (LIGO) have  become the first strong evidence of the existence of black holes in the universe\cite{GW1}. In 2019, the pictures of a supermassive black hole in the center of the giant elliptical galaxy M87 released by the international collaboration of Event Horizon Telescope (EHT) directly confirmed the existence of black hole\cite{EHT1,EHT2,EHT3,EHT4,EHT5,EHT6}.
According to the picture obtained by EHT, there is a dark area with obviously insufficient observation intensity inside the bright ring, which is called the black hole shadow and the bright ring is a photon sphere.
The light ray from cosmic accretion materials is absorbed by black hole due to its strong gravitational field and cannot reach the distant observers, thus forming the shadow of black hole\cite{Cunha:2018acu}. In 1966, Synge proposed the theoretical condition that photons can escape from the strong gravitational field of black hole\cite{Lens1}, which indicated that the shadow contour of static spherically symmetric black hole is a standard circle\cite{Lens2,Shape1,Shape2}. After that, Bardeen obtained that the radius of Schwarzschild black hole is $r=5.2M$, in which $M$ is the mass of black hole\cite{Bar}. On the contrary, the shadow of rotating black holes show a D-shaped shape, which is related to the spin parameter\cite{Shape4,Shape5,Shape6,Shape7,Shape8}. In the various gravity backgrounds, apart from the study of the shape and size of shadows,  the black hole shadows surrounded by different accretion models and their observational characteristics  have also been studied, and references therein\cite{spherical1,spherical3,spherical4,spherical5,spherical6,spherical7,spherical8,spherical9,spherical10,spherical11,SPL,thin1,thin2,GMY1,GMY2,GMY4}.

The shadow of a black hole contains a lot of information. The study of shadow not only enables us to comprehend the geometric structure of spacetime, but also helps us exploring various gravity models more deeply. However, the above research on black hole shadows is based on the geometric optics, i.e., the famous ray-tracing method.
Therefore, in the framework of the wave optics, the holographic image of AdS black hole in the bulk was constructed when the wave emitted by the source at the boundary of AdS enters the bulk and then propagate in the bulk by considering the AdS/CFT correspondence \cite{Hashimoto:2019jmw,Hashimoto:2018okj}. In particular, the Einstein ring can be clearly observed in the framework of holographic, and the size of the ring is consistent with the size of the black hole photon sphere obtained from geometric optics.
On later, the Einstein ring structure of the lens response of the complex scalar field has been studied in the background of a charged AdS black hole, and the results show that the radius of the Einstein ring does not change with the chemical potential, but is very dependent on the change of temperature \cite{Liu:2022cev}. In addition, the asymptotically AdS black hole dual to a superconductor is imaged in \cite{Kaku:2021xqp}, and the authors further  investigate the effect of the charged scalar condensate on the image, where the Maxwell field is considered. Indeed, the holographic images can be used as an effective method to test the existence of the gravitational dual for given materials, and it further allows us to comprehend the configuration of the dual black hole.

It is well known that, the Gauss-Bonnet gravity is a well-modified gravity model, and the equation of motion has no higher derivative than the second order. In four-dimensional spacetime, the Gauss-Bonnet term in the Lagrangian is topologically invariant and thus does not contribute to the dynamics, but it contributes to the dynamics of gravitational field in high-dimensional ($d>4$) cases \cite{Cai:2001dz}. Recently, Glavan and Lin show a Gauss-Bonnet modified gravity in four dimension with adjusting the Gauss-Bonnet (GB) coupling constant to $\alpha \rightarrow \alpha / (d-4)$, where $d$ take the limit $d \rightarrow 4$ \cite{Glavan:2019inb}. However, their works  does not lead to a well-defined way with the initial regularization scheme\cite{Hennigar:2020lsl,Shu:2020cjw}. To overcome this problem, the author found a well defined and consistent theory by breaking the temporal diffeomorphism property of the curved spacetime in \cite{Aoki:2020iwm}. In the framework of Gauss-Bonnet gravity, the influence of Gauss-Bonnet constant on black hole shadow and photon sphere dynamics is studied \cite{spherical2,Konoplya:2020bxa}. As previously mentioned, those works are based on the method of geometrical optics. In view of this,  the main purpose of this work is to use the wave optics method to study the holographic image of the Gauss-Bonnet AdS black hole\cite{Hennigar:2020lsl}.
Following the idea in \cite{Hashimoto:2019jmw,Hashimoto:2018okj}, we intend to take an oscillatory Gaussian wave source $\mathcal{J}_o$ on one side of the AdS boundary, and scalar waves generated by the source can propagate in the bulk. After the bulk of scalar wave reach the other side of AdS boundary, the corresponding response will be generated. By using a special optical system, we can convert the extracted response function $\langle \mathcal{O} \rangle$ into the holographic image that can be seen on a screen. Here\footnote{The schematic diagram of the working principle is shown in Fig.1.}, the ($2+1$)-dimensional boundary CFT on the $2$-sphere $S^2$  is naturally dual to a black hole in the global $AdS_4$ spacetime, or the massless volume scalar field in the spacetime. Hence, we can  explore whether the dependence of the holographic image characteristics of Gauss-Bonnet AdS black hole is reflected in the Gauss-Bonnet coupling constant or wave sources, whereby acquire a greater depth of understanding of the gravitational dual for given materials.

\begin{figure}[!h]
\makeatletter
\renewcommand{\@thesubfigure}{\hskip\subfiglabelskip}
\makeatother
\centering 
\subfigure[]{
\setcounter{subfigure}{0}\subfigure[]{\includegraphics[width=0.4\textwidth]{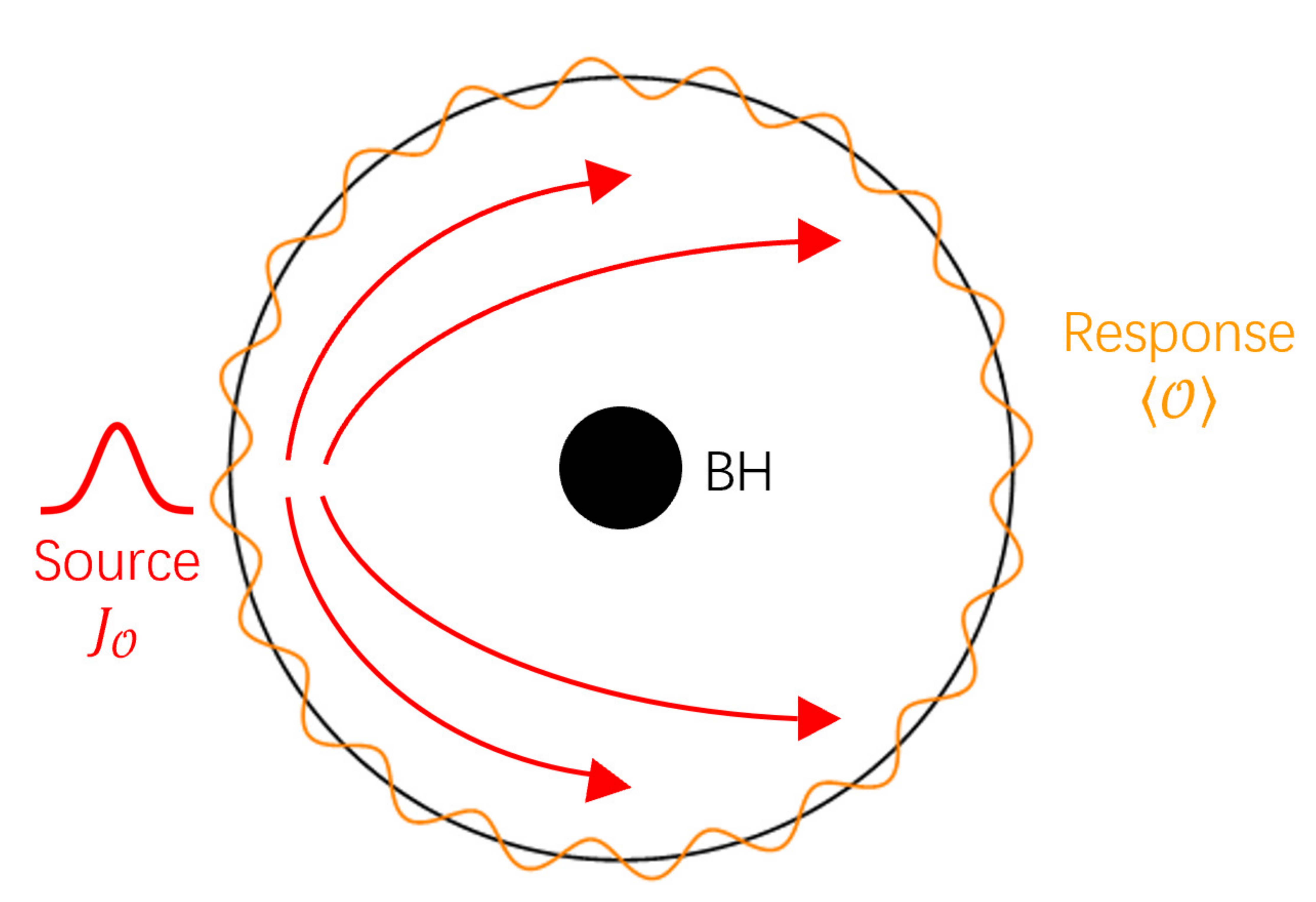}}}
\caption{\label{figres11} The schematic of imaging a dual black hole. }
\end{figure}

The remainders of the present paper are outlined as follows. In  Section \ref{sec2},  we shall briefly introduce Gauss-Bonny gravity, and extract the response function at the north pole when the Gaussian wave source is located at the South Pole of the AdS boundary. In  Section \ref{sec3}, we will introduce a special optical system consisting of a convex lens and a spherical screen. With the help of this imaging system, the holographic image of  Gauss-Bonnet AdS black hole is constructed through the obtained response function. In this way, we can study the effects of state parameters on the holographic image of the black hole.
Section \ref{Concl} ends up with a brief discussion and conclusion.

\section{Scalar field and response function in a Gauss-Bonnet AdS black hole }
\label{sec2}
Firstly, we will introduce the Gauss-Bonnet AdS black hole in this section. In general, for a $d$-dimensional spacetime with a negative cosmological constant, the action in Gauss-Bonnet gravity reads
\begin{align}\label{eq1}
\mathcal{S}= \frac{1}{16 \pi} \int d^d x \left(R+\frac{(d-1)(d-2)}{l^2} +\frac{\alpha}{d-4} L_{GB} -F_{\mu\nu}F^{\mu\nu}  \right),
\end{align}
and
\begin{align}\label{eq2}
L_{GB}= R_{\mu\nu\rho\sigma}R^{\mu\nu\rho\sigma}-4R_{\mu\nu}R^{\mu\nu}+R^2,
\end{align}
where $l$ is the AdS radius which is related to the cosmological constant, $F_{\mu\nu}$ and $\alpha$ are the Maxwell tensor and the GB coupling parameter, respectively. By rescaling the Gauss-Bonnet coupling parameter $\alpha\rightarrow\alpha/(d-4)$ and taking the limit $d-4$, one can obtain a 4-dimensional nontrivial black hole solution, which is
\begin{align}\label{eq3}
ds^2= -F(r)dt^2 + \frac{1}{F(r)} dr^2 + r^2 d\theta^2 +r^2 {\sin \theta}^2 d\varphi^2,
\end{align}
with
\begin{align}\label{eq4}
F(r)= r^2 f(r) = 1 + \frac{r^2}{2\alpha}\left( 1- \sqrt{1+4\alpha \left(\frac{2M}{r^3} -\frac{1}{l^2} \right)} \right),
\end{align}
By using a new definition $u=1/r$, one can use the new coordinate ${(t,u,\theta,\varphi)}$ to reexpress the metric function $f(r)$. In the Eddington ingoing coordinate, i.e., $v = t+u_* = t- \int \frac{du}{f(u)}$, the metric function can be further rewritten as following, which is
\begin{align}\label{eq7}
ds^2= \frac{1}{u^2}\left(-f(u)dv^2 -2dudv + d\theta^2 + {\sin \theta}^2 d\varphi^2 \right),
\end{align}
where,
\begin{align}\label{eq8}
f(u)= 1 + \frac{1}{2\alpha u^2}\left( 1 - \sqrt{1-\alpha \left(4 - 4 u^2 u_h\left(\frac{1}{u_h^4} +\frac{1}{u_h^2} + \alpha  \right) \right)} \right),
\end{align}
where $u_h = 1/r_h$, $r_h$ is the event horizon of the black hole which can be obtained with $f(r)=0$, and the gauge symmetry has been considered. For a massless particle in the scalar field, the Klein-Gordon equation is
\begin{align}\label{eq10}
\Box \Phi(v,u,\theta,\varphi)=0,
\end{align}
For the spacetime (\ref{eq7}), we have
\begin{align}\label{eq11}
&u^2 f(u)\partial_u \partial_u \Phi + \left[u^2 f'(u) - 2u f(u) \right] \partial_u \Phi-2 u^2\partial_v \partial_u \Phi
 + 2u \partial_v\Phi + u^2 D^2_S \Phi =0,
\end{align}
where $f'(u)=\partial_u f(u)$. The asymptotic solution of Eq.(\ref{eq11}) near the AdS boundary($z\rightarrow0$) reads \cite{Liu:2022cev}
\begin{align}\label{eq12}
\Phi(v,u,\theta,\varphi) = \mathcal{J}_{O}(v,\theta,\varphi) + u \partial_v \mathcal{J}_O(v,\theta,\varphi) +\frac{1}{2} u^2 D_S^2\mathcal{J}_O(v,\theta,\varphi)+ u^3 \langle O \rangle + \mathcal O (u^4).
\end{align}
Here, $D_S^2$ represents the scalar Laplacian on unit $S^2$. According to the AdS/CFT dictionary, it is obvious that the $\mathcal{J}_{O}(v,\theta,\varphi)$ and $\langle O \rangle$ are the external scalar source and corresponding response function in the dual CFT, respectively. In this paper, we employ the monochromatic and axissymmetric Gussian wave packet source as the external scalar source, and fixed it at the south pole of the AdS boundary as the source. In this sense, we have,
\begin{align}\label{eq13}
\mathcal{J}_O(v,\theta) = e^{i \omega v} \cdot \frac{1}{2 \pi \sigma^2  } \cdot \exp\left[ -\frac{(\pi - \theta)^2}{2 \sigma^2} \right] = e^{i \omega v} \cdot \sum_{l=0}^{\infty} C_{l0} Y_{l0}(\theta),
\end{align}
with
\begin{align}\label{eq14}
C_{l0} = (-1)^l \sqrt{\frac{l+1/2}{2 \pi}} \exp \left[ -\frac{1}{2} (l+1/2)^2 \sigma^2 \right].
\end{align}
In $C_{l0}$, $\sigma$ and $Y_{l0}$ are the width of the wave produced by the Gussian source and the spherical harmonics function, respectively. And, we only consider the case $\sigma \ll \pi$ because the tiny value of Gaussian tail can be neglected. Considering the symmetry of Eq.(\ref{eq3}), one can further decompose the function $\Phi(v,z,\theta,\varphi)$ as,
\begin{align}\label{eq15}
\Phi(v,u,\theta,\varphi) = e^{i \omega v} \cdot \sum_{l=0}^{\infty} \sum_{m=-l}^{l} c_{l0} U_l(u)  Y_{l0}(\theta,\varphi).
\end{align}
And, the response function reads
\begin{align}\label{eq16}
\langle O \rangle = e^{i \omega v} \sum_{l} \langle O \rangle_l Y_{l0}(\theta)
\end{align}.
With the aid of Eq.(\ref{eq15}), we have
\begin{align}\label{eq17}
u^2 f(u) U''_l +\left[u^2 f'(u) -2 u f(u) + 2 i \omega u^2 \right] U'_l + \left[ - 2 i \omega u - l(l+1) u^2 \right] U_l =0.
\end{align}
And the asymptotic behaviour of $U_l$ can be written as
\begin{equation}\label{eq18}
 \lim_{z \rightarrow 0} U_l= 1 - i \omega u + \frac{1}{2}\left[-l(l+1) \right]u^2 +{\langle O \rangle_l} u^3 + \mathcal O (u^4)
\end{equation}.
For Eq.(\ref{eq15}), it is obvious that there are two boundary conditions for the function $U_l$. One is the horizon boundary condition, which is
\begin{equation}\label{eq118}
\left[u_h^2 f'(u_h) + i 2 \omega u_h^2 \right]U'_l - \left[ 2 i\omega u_h +l(l+1)u_h^2 \right]U_l=0
\end{equation}
at the event horizon $u=u_h$. And, another is AdS boundary condition, which is $U_l(0)=1$ at the AdS boundary. Combined with those conditions, one can solve the Eq.(\ref{eq17}) and obtained the function $U_l$ by employing the psuedo-spectral method\cite{Liu:2022cev}. Then, the total response function $\langle O \rangle$ can be found with the aid of the Eqs.(\ref{eq18}) and (\ref{eq16}). Here, we take some proper values of black hole and lens parameters as examples to clearly show the absolute amplitude of $\langle O \rangle$, which can be seen in Fig.2. In Fig.2 (a), the values of GB coupling parameter change, while the values of other relevant state parameters do not change, where $M=1$, $r_h=1$, $\sigma=0.05$, and $\omega=75$. {In Fig.2 (b), the value of GB coupling parameter is fixed $\alpha=0.15$, while the value of other relevant state parameters changes, in which $r_h=0.6, \sigma=0.05,\omega=75$(the gray dashed), $r_h=0.8, \sigma=0.05,\omega=75$(the blue dashed), $r_h=0.6, \sigma=0.06,\omega=75$(the black dashed) and $r_h=0.6, \sigma=0.05,\omega=80$(the green dashed).}
\begin{figure}[!h]
\makeatletter
\renewcommand{\@thesubfigure}{\hskip\subfiglabelskip}
\makeatother
\centering 
\subfigure[]{
\setcounter{subfigure}{0}\subfigure[($a$)]{\includegraphics[width=0.4\textwidth]{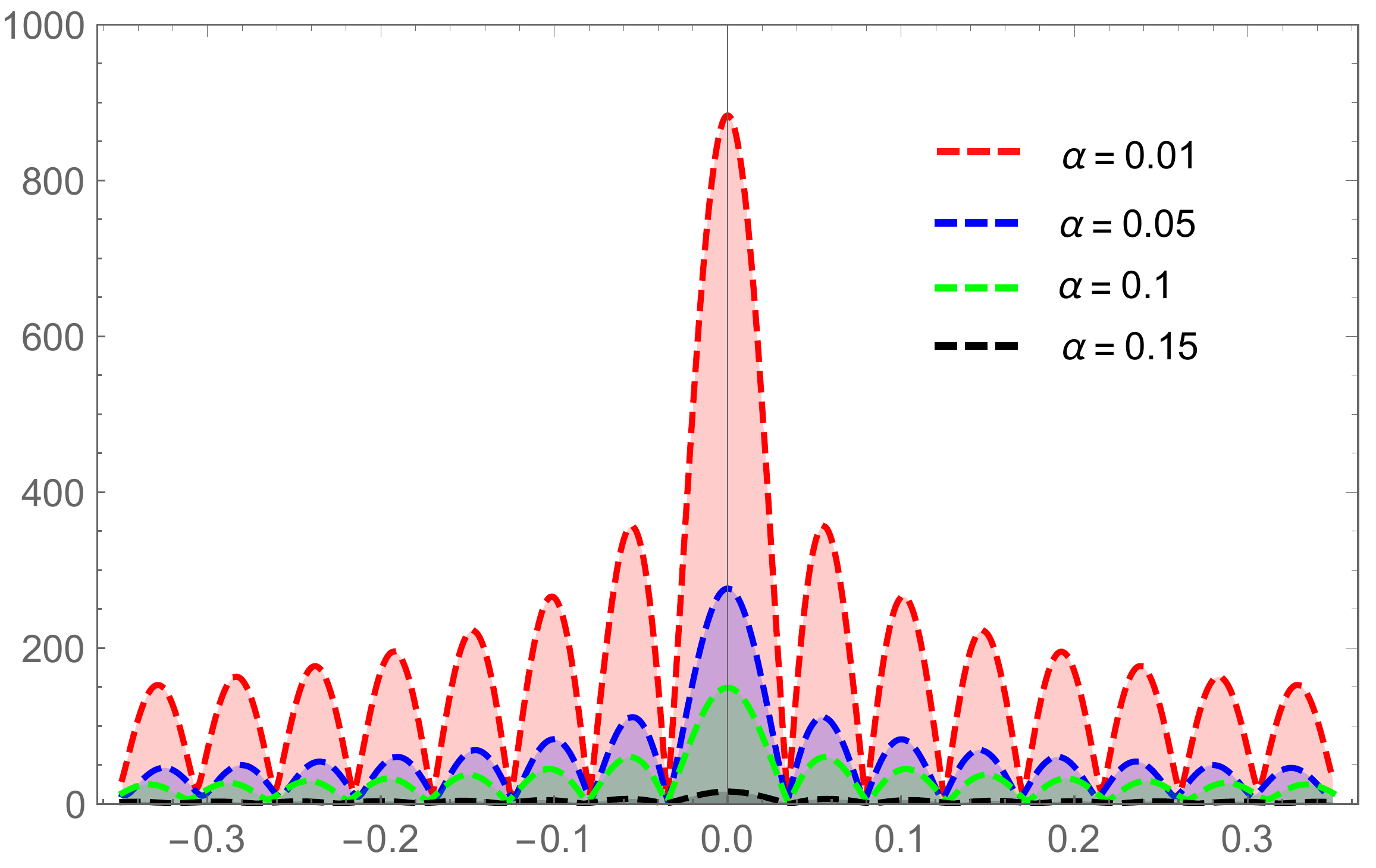}}
\subfigure[($b$)]{\includegraphics[width=0.4\textwidth]{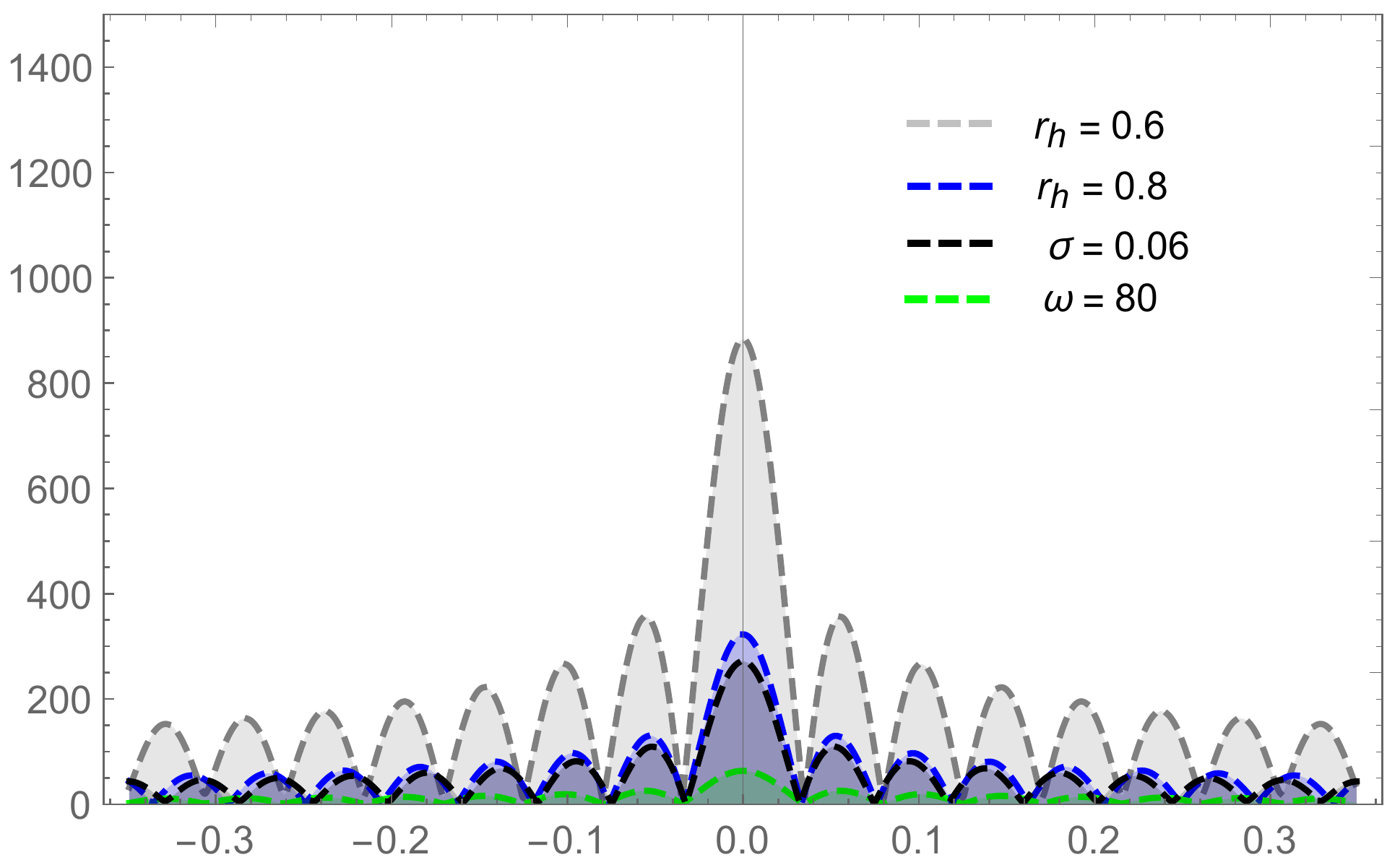}}}
\caption{\label{figres}  The absolute amplitude of total response function for different values of black hole and lens parameters. }
\end{figure}

{From Fig.2, one can obviously observes the diffraction pattern as the scalar wave propagates in the bulk and is diffracted by the black hole. {More importantly, it shows that the absolute amplitude of total response function increases with the decrease of the GB coupling parameter $\alpha$ and  event horizon of black hole $r_h$. In addition, we find that the width and frequency of Gaussian source also diminished the absolute amplitude.} In other words, the total response function depends closely on the Gaussian source and the spacetime geometry. Therefore, if this response function can be transformed as the observed images, it can be regarded as a useful tool to reflect the feature of the spacetime geometry. To achieve this goal, a special optical system is required, which will be described in detail in the next section.}

\section{Holographic rings of Gauss-Bonnet AdS black hole }
\label{sec3}
\begin{figure}[!h]
\makeatletter
\renewcommand{\@thesubfigure}{\hskip\subfiglabelskip}
\makeatother
\centering 
\subfigure[]{
\setcounter{subfigure}{0}\subfigure[($a$)]{\includegraphics[width=0.33\textwidth]{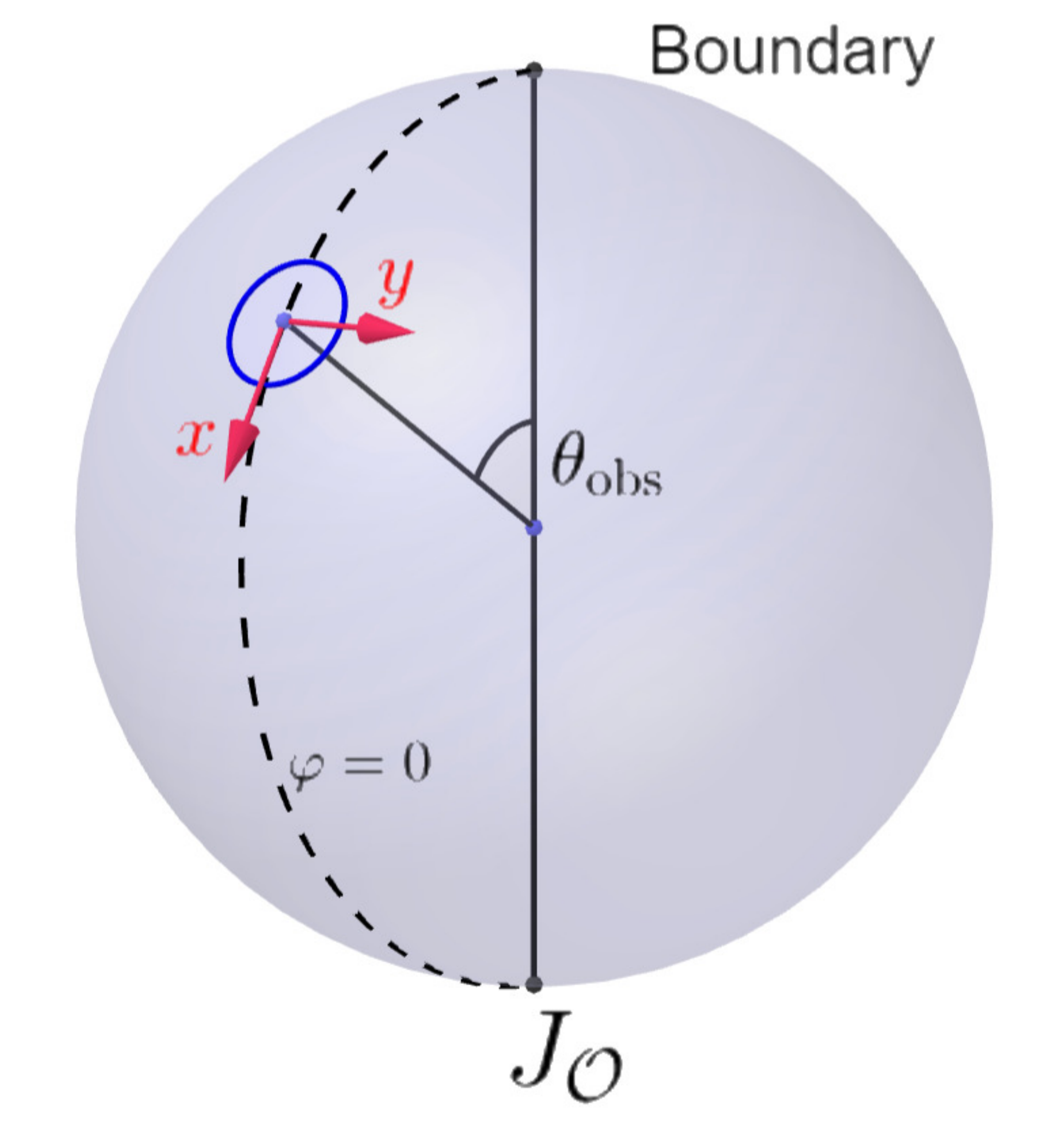}}
\subfigure[($b$)]{\includegraphics[width=0.4\textwidth]{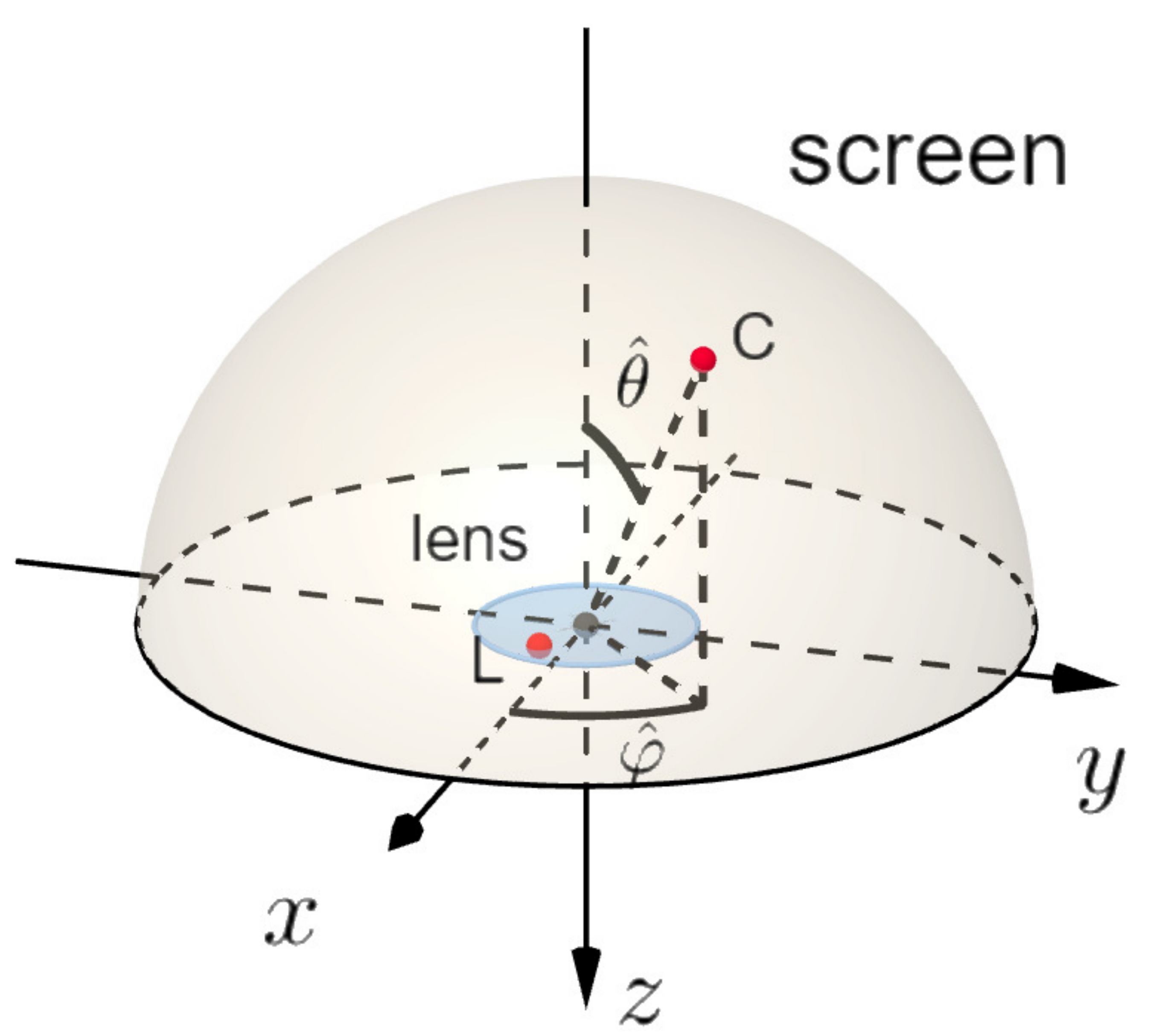}}}
\caption{\label{figre2}  (a): The region on the boundary; (b): the structure of optical system. }
\end{figure}

After obtaining the response function, we will use it to directly observe the black hole images in this section. At this moment, we need to use a special optical system, which is composed of an extremely thin convex lens and spherical screen. By considering that the observation area on the  AdS boundary  is a very small range, where the observation center is ($\theta_{obs}, 0$), as shown in the blue circle in Fig.3 (a). The convex lens is placed in the observation range, that is, its coordinate position is $\vec{x}=(x, y, 0)$, and the coordinates of the spherical screen is  $\vec{x}_S=(x_S , y_S , z_S )$. And, the focal length $f$ of  the infinitely thin lens is much larger than the size of it, i.e., $f\gg d$. At the observed point, the response function as the plane wave $\Psi_p(\vec{x})$ will be converted as the transmitted wave $\Psi_s(\vec{x})$ by using the lens.  This wave as the spherical wave will convert to the observed wave $\Psi_{sc}(\vec{x}_s)$ when it reached to the screen, which is shown in Fig.3 (b).
In this case, we have
\begin{equation}\label{eq19}
\Psi_{sc}(\vec{x}_s) = \int_{|\vec{x}|<d} dx^2 \Psi_s(\vec{x}) e^{- i \omega L} = \int_{|\vec{x}|<d} dx^2 e^{- i \omega \frac{|\vec{x}|^2}{2f}} \Psi_p(\vec{x}) e^{- i \omega L},
\end{equation}
where $f^2 = x_S^2 + y_S^2+ z_S^2 $, and the term $L$ is the distance between $\vec{x}$ and $\vec{x}_S$. By considering $L = \sqrt{(x_s-x)^2+(y_s-y)+z_s^2}\simeq f - \frac{\vec{x}_s \cdot \vec{x}}{f} + \frac{|\vec{x}|^2}{2f}$ and the Fresnel approximation $f\gg|\vec{x}|$, one can get
\begin{equation}\label{eq21}
\Psi_{sc}(\vec{x}_s) \propto \int_{|\vec{x}|<d} dx^2 \Psi_p(\vec{x}) \varpi(\vec{x}) e^{-\frac{i \omega }{f} \vec{x}\cdot \vec{x}_s}
\end{equation}
with the window function reads
\begin{align} \label{Eq22}
  { \varpi(\vec{x})}\equiv
    \begin{cases}
   \text{1}, \quad \quad 0 \leq \mid \vec{x}\mid \leq d\\
     \text{0}, \quad \quad  \mid \vec{x}\mid > d
    \end{cases}.
\end{align}
To find Eq.(\ref{eq21}), here we have used the Taylor expansion and some properly approximations. And from it, it shows that the observed wave on the screen connect with the incident wave by the Fourier transformation. As long as identified the respond function as the function on the lens $\Psi_p=\langle O \rangle$, we will capture the images of the dual black hole on the screen by using Eq.(\ref{eq21}). When the observer located at different positions of AdS boundary, change the relevant parameters ($\alpha$ and $r_h$) of the spacetime system, and the holographic Einstein images will be obtained, which have been presented in Figs.\ref{1fig666} and \ref{2fig666}\footnote{Here, the vertical line is $ys/f$, and horizontal axis is $xs/f$, and the range of them belongs to $(-1.5,1.5)$ which also used in later similar figures.}.
\begin{figure}[!h]
\makeatletter
\renewcommand{\@thesubfigure}{\hskip\subfiglabelskip}
\makeatother
\centering 
\subfigure[(a): $\alpha = -0.05, r_h=0.6$]{
\setcounter{subfigure}{0}\subfigure[$\theta_{obs}=0^\circ$]{\includegraphics[width=.22\textwidth]{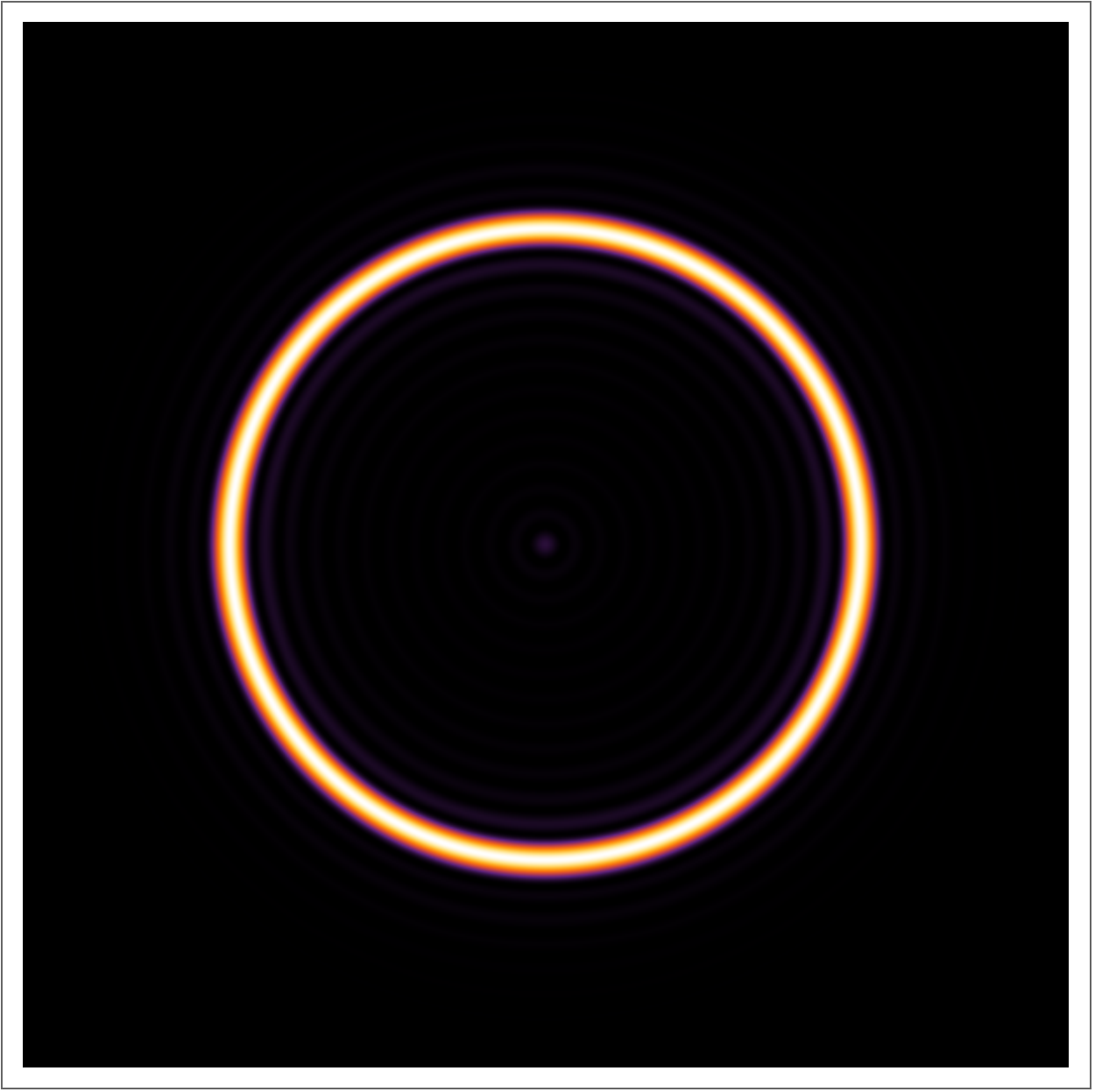}}
\subfigure[$\theta_{obs}=30^\circ$]{\includegraphics[width=.22\textwidth]{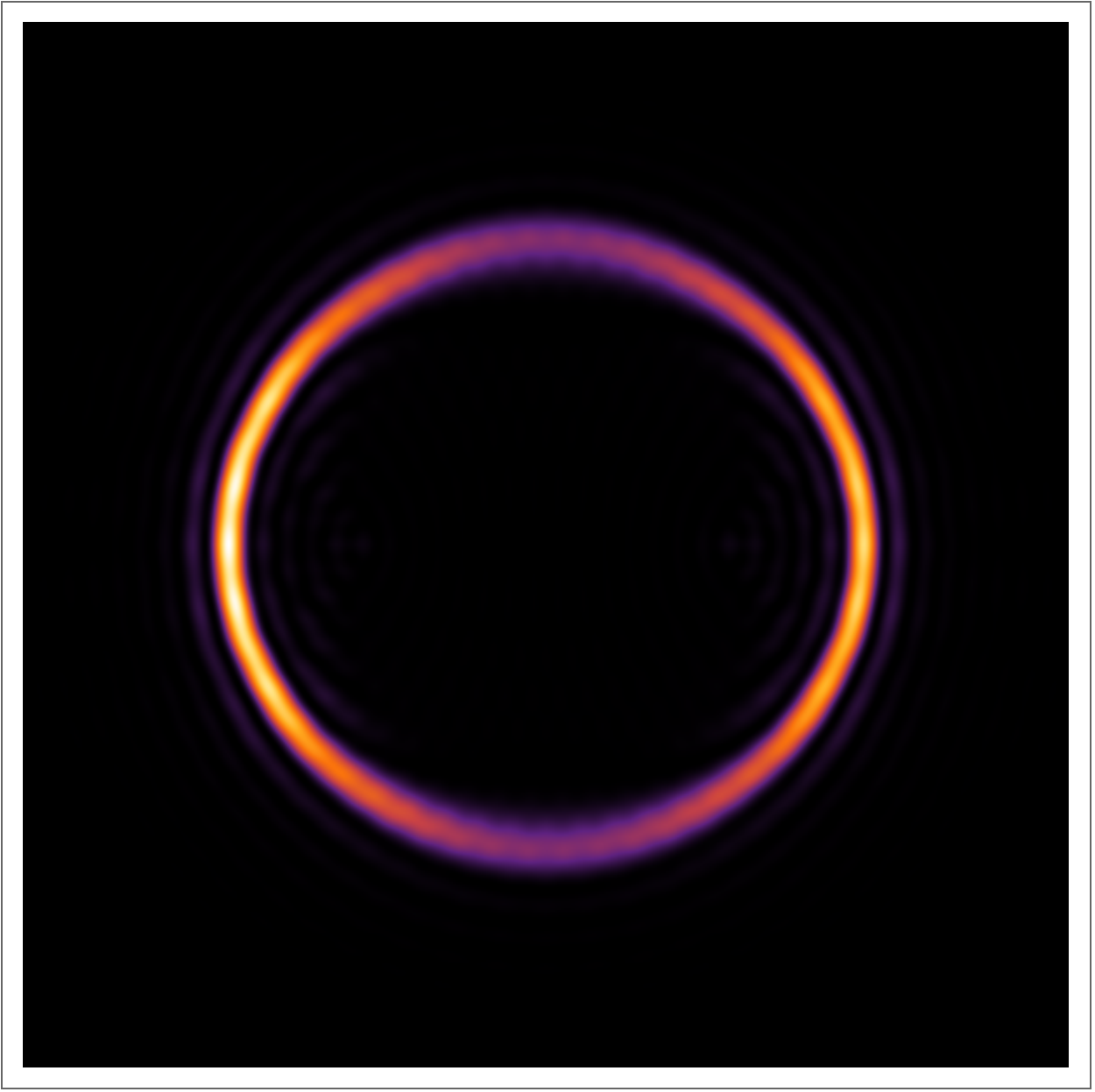}}
\subfigure[$\theta_{obs}=60^\circ$]{\includegraphics[width=.22\textwidth]{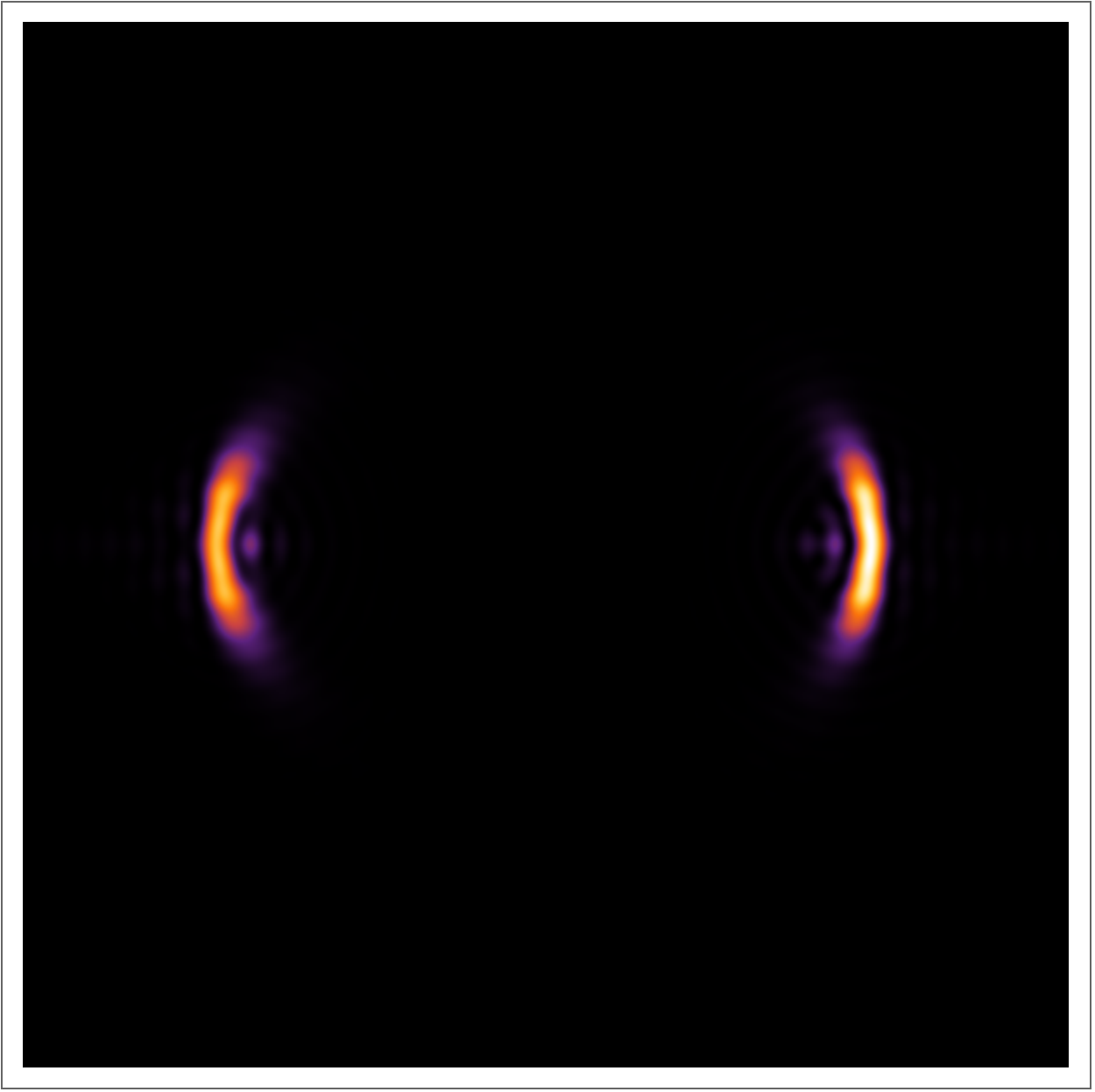}}
\subfigure[$\theta_{obs}=90^\circ$]{\includegraphics[width=.22\textwidth]{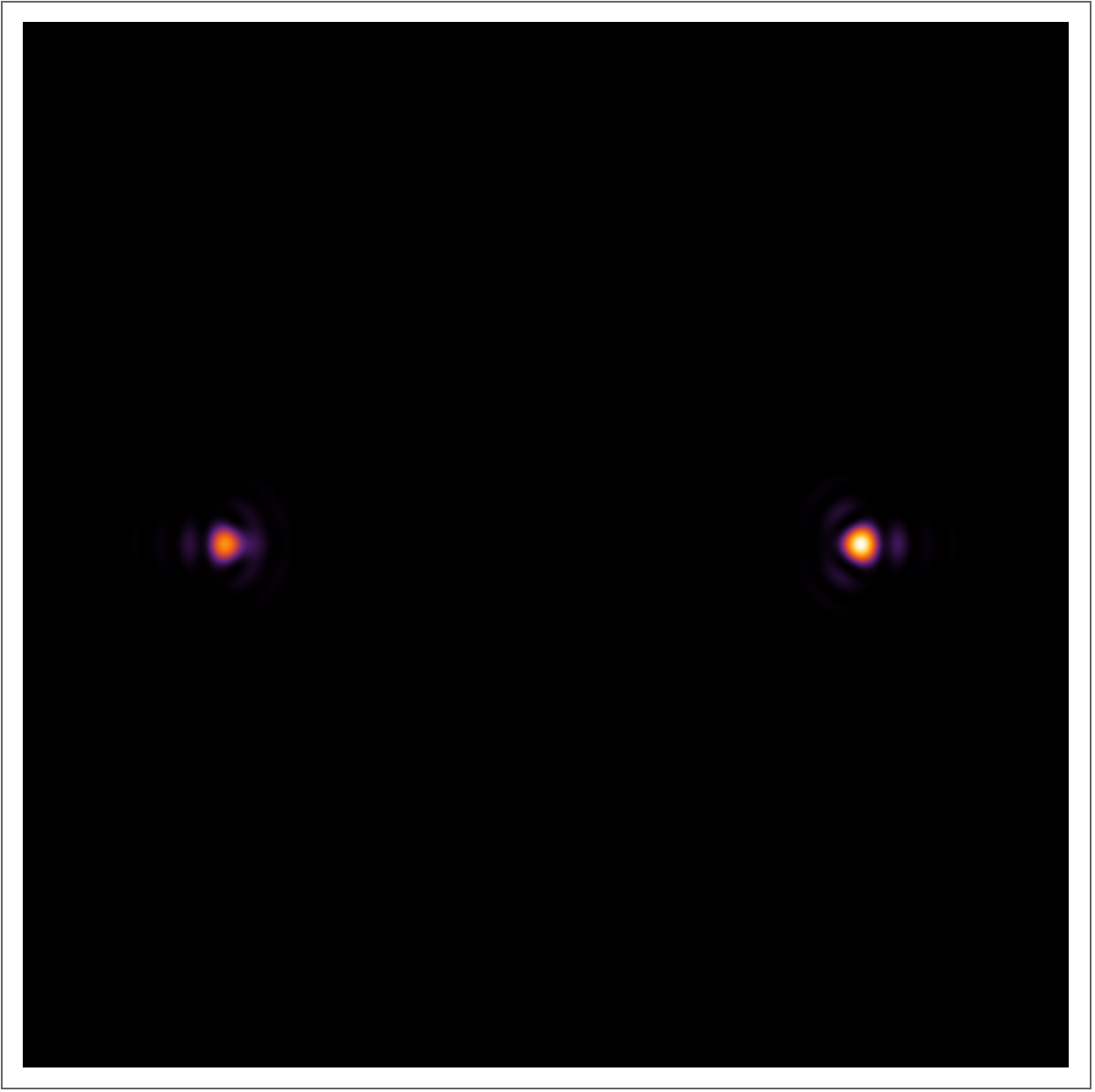}}}
\end{figure}

\begin{figure}[!h]
\makeatletter
\renewcommand{\@thesubfigure}{\hskip\subfiglabelskip}
\makeatother
\centering 
\subfigure[(b): $\alpha = 0.05, r_h=0.6$]{
\setcounter{subfigure}{0}\subfigure[$\theta_{obs}=0^\circ$]{\includegraphics[width=.22\textwidth]{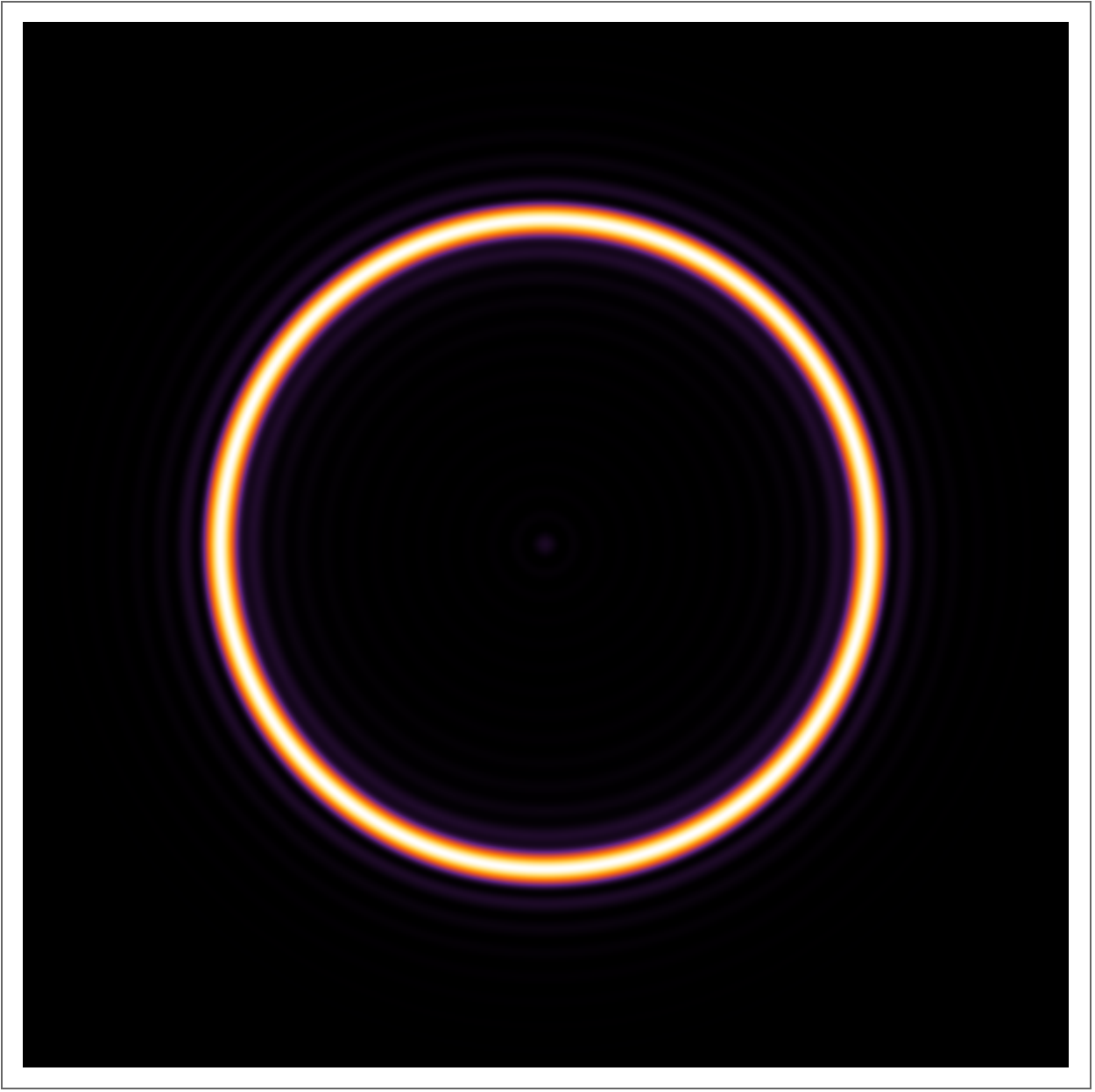}}
\subfigure[$\theta_{obs}=30^\circ$]{\includegraphics[width=.22\textwidth]{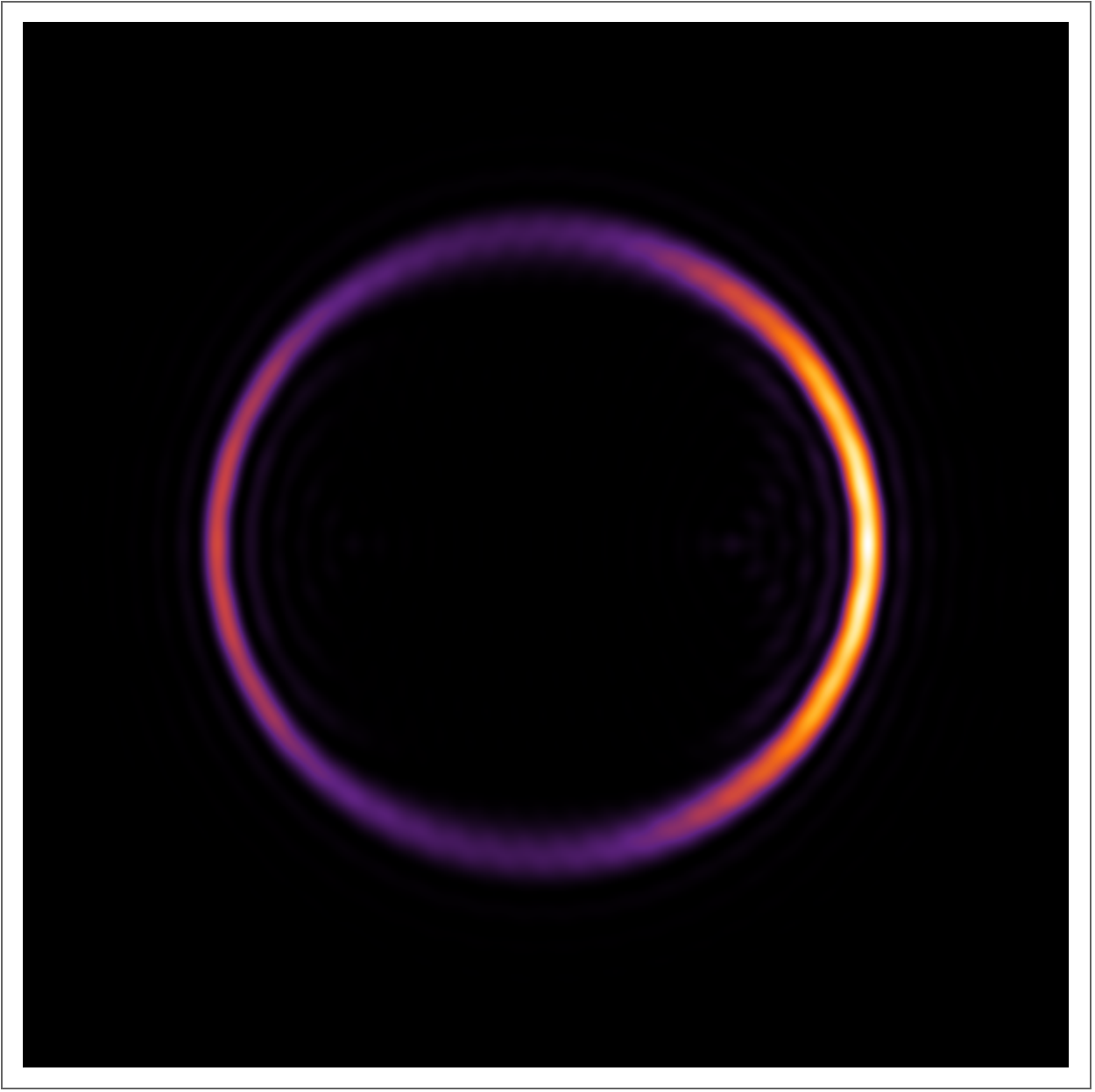}}
\subfigure[$\theta_{obs}=60^\circ$]{\includegraphics[width=.22\textwidth]{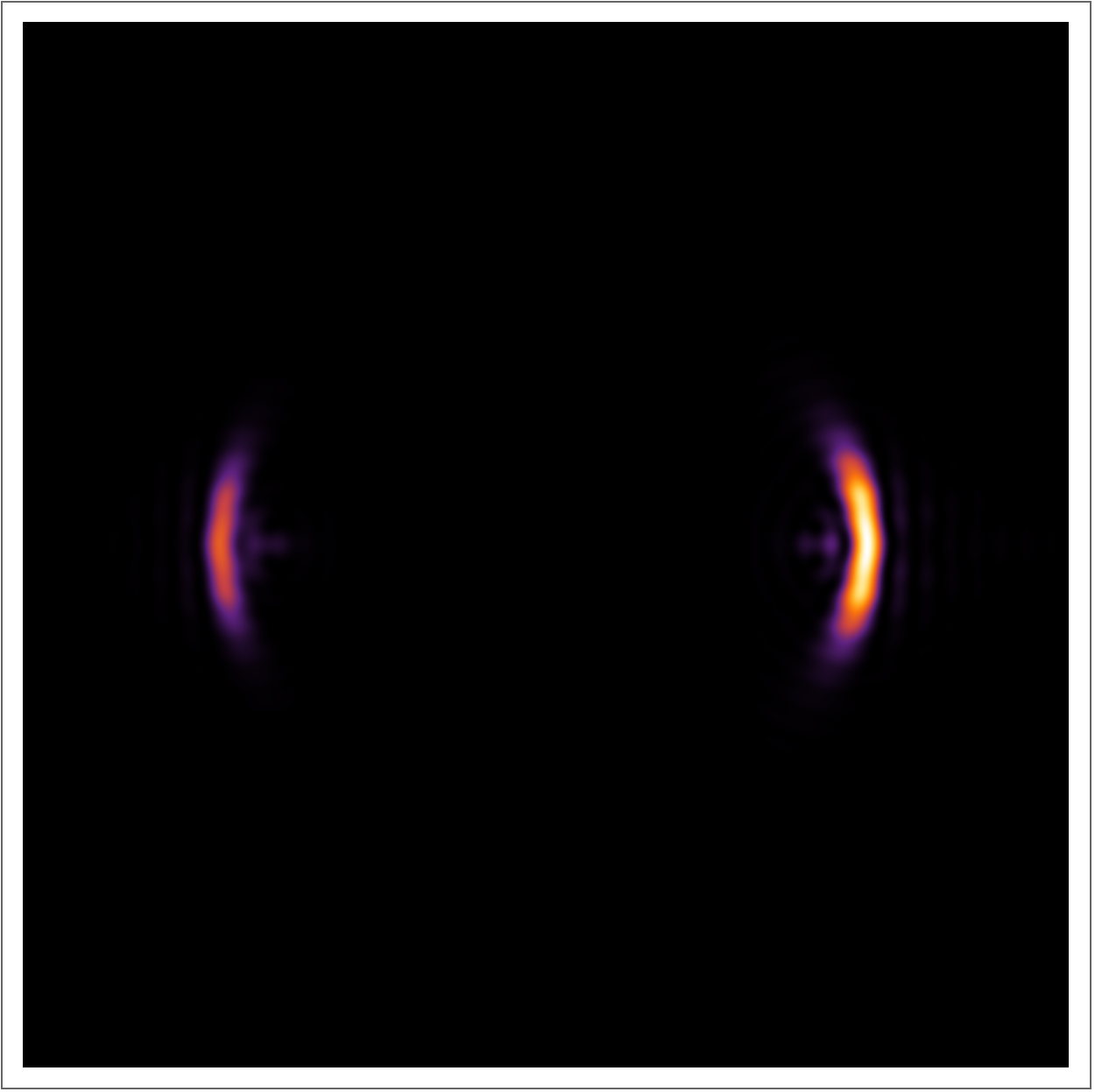}}
\subfigure[$\theta_{obs}=90^\circ$]{\includegraphics[width=.22\textwidth]{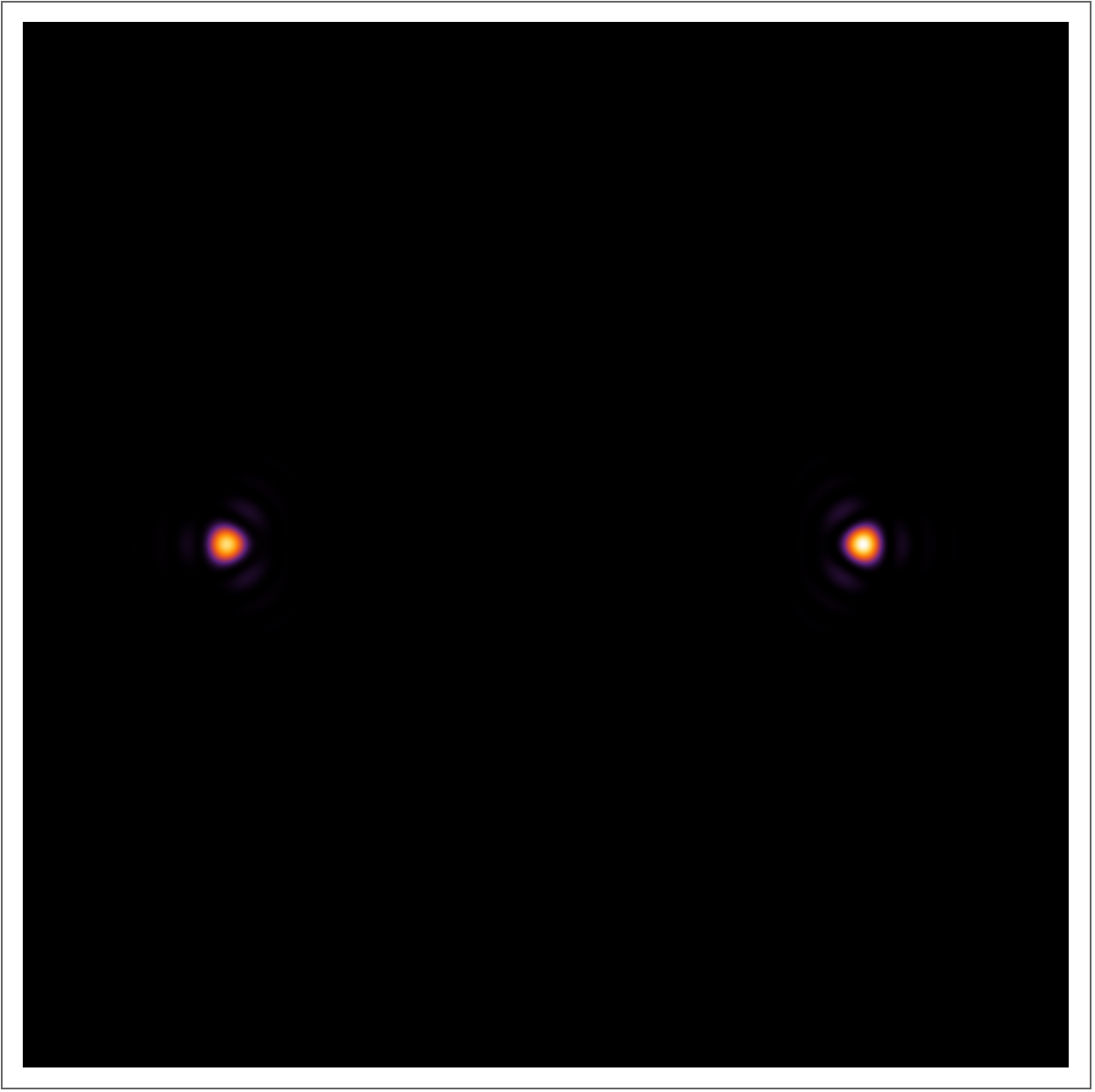}}}
\end{figure}
\begin{figure}[!h]
\makeatletter
\renewcommand{\@thesubfigure}{\hskip\subfiglabelskip}
\makeatother
\centering 
\subfigure[$(c): \alpha = 0.1, r_h=0.6$]{
\setcounter{subfigure}{4}\subfigure[$\theta_{obs}=0^\circ$]{\includegraphics[width=.22\textwidth]{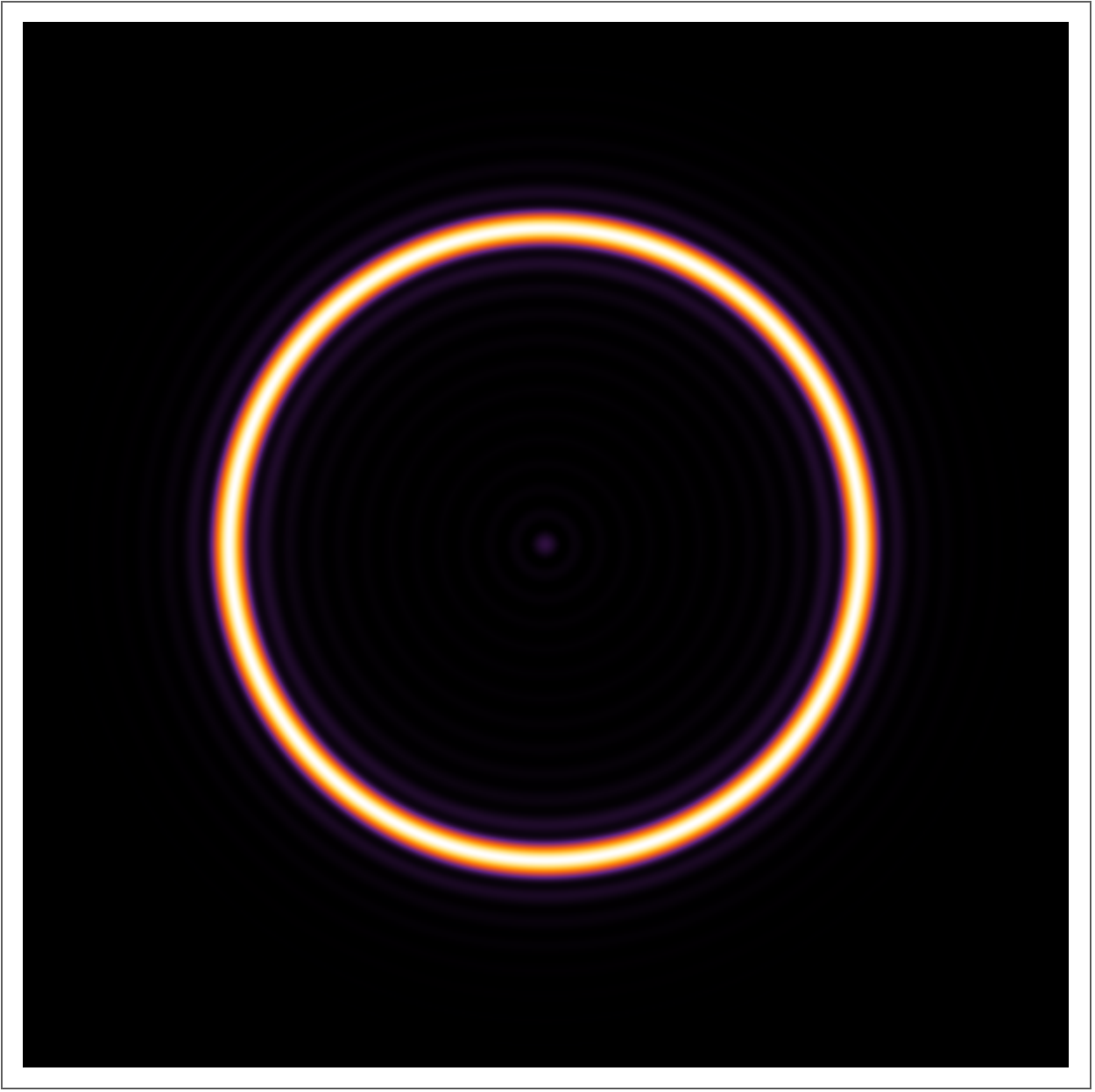}}
\subfigure[$\theta_{obs}=30^\circ$]{\includegraphics[width=.22\textwidth]{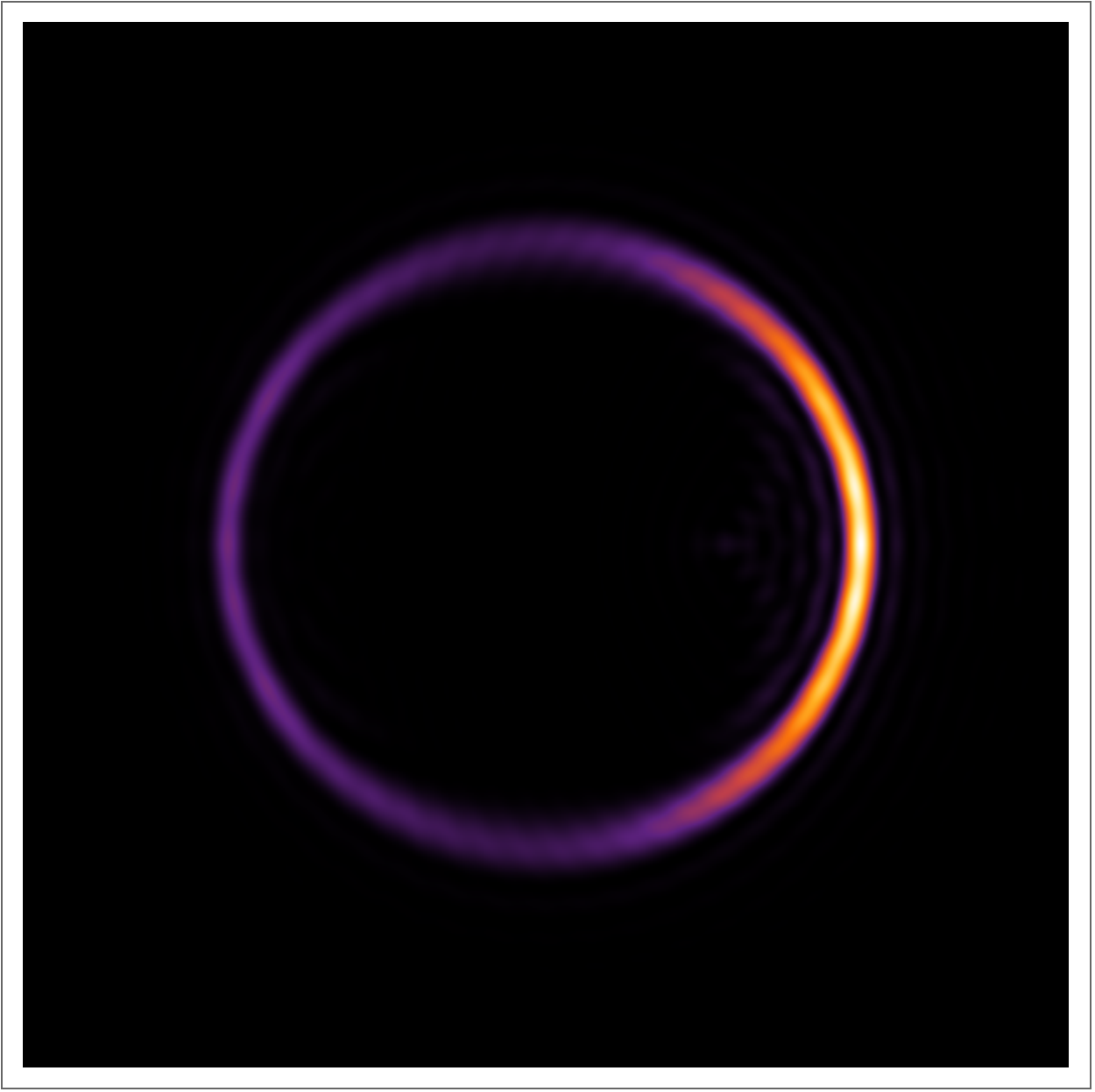}}
\subfigure[$\theta_{obs}=60^\circ$]{\includegraphics[width=.22\textwidth]{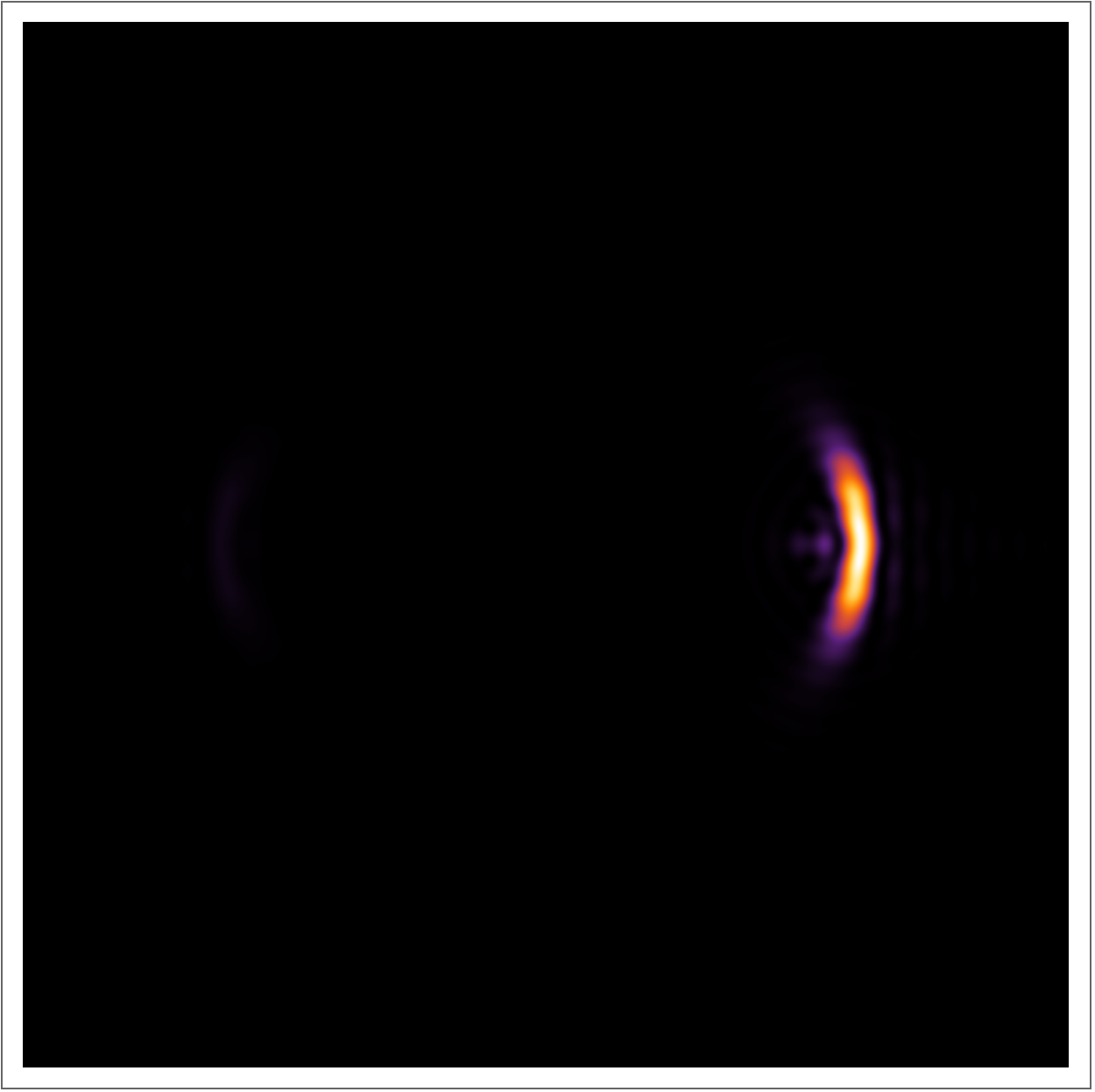}}
\subfigure[$\theta_{obs}=90^\circ$]{\includegraphics[width=.22\textwidth]{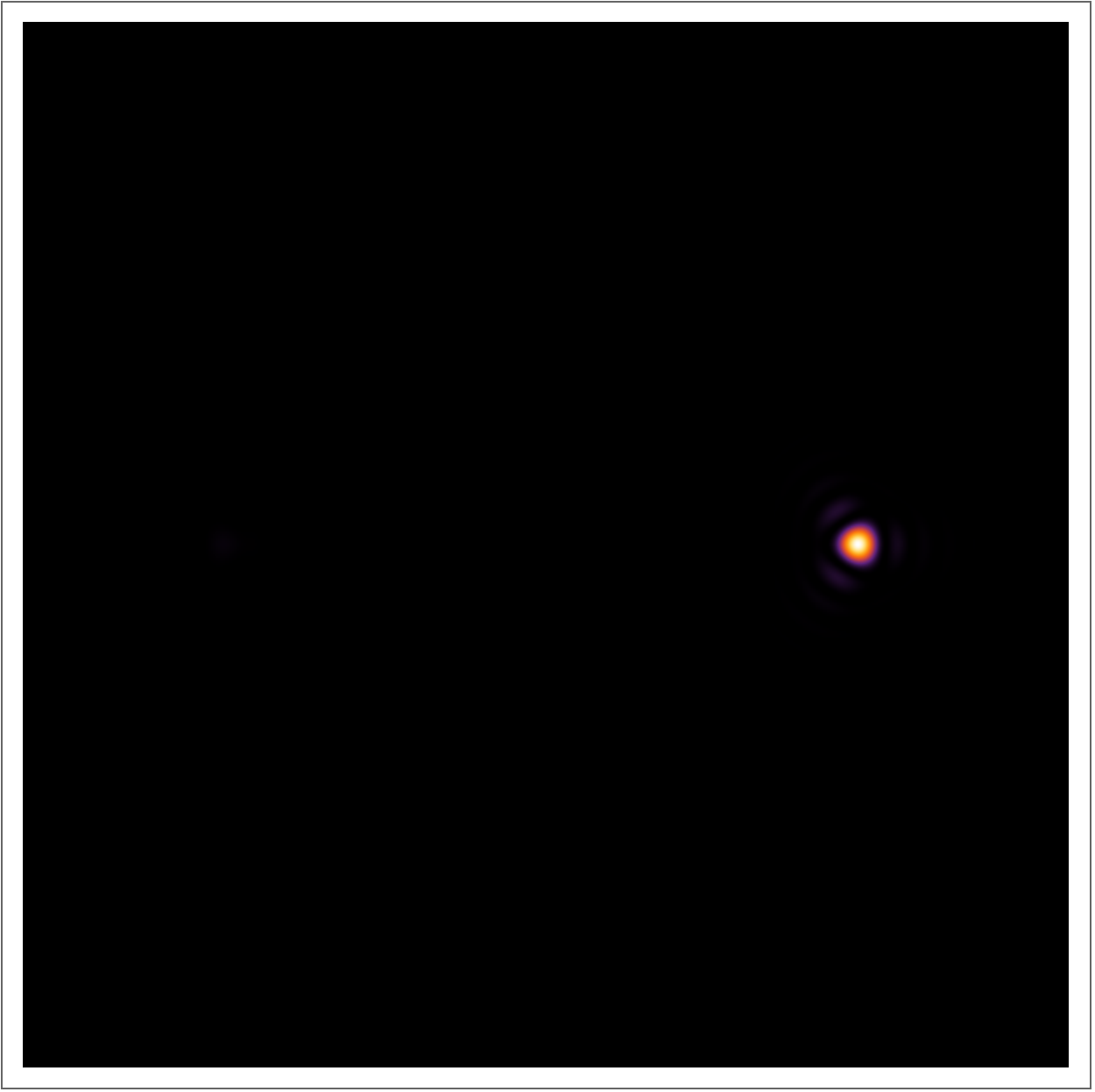}}}
\caption{\label{1fig666} The holographic images for $d=0.6, \sigma=0.05, \omega=75.$}
\end{figure}

When the observer located at the position $\theta=0^\circ$, i.e., the observated location is the north pole of the AdS boundary, it can be seen that a series of axisymmetric concentric rings appear in the image, and one of them is particularly bright. At the center of the rings, there is a bright spot which called as the Poisson$-$like spot.
As the observed position fixed to $\theta=30^\circ$, the bright ring changed into a luminosity-deformed ring, which is instead of a strict axisymmetric ring.
And at $\theta=60^\circ$, two bright light arcs appeared, rather than a ring.
In particular, this ring evolved into two light points finally in the subfigure(a) in Fig.\ref{1fig666}, which correspond to the clockwise and anticlockwise light rays respectively from the viewpoint of geometrical optics.

With the change of GB coupling constant $\alpha$ in the system, the holographic images is different. When the observer located at the position $\theta=0^\circ$, one can find that the size and brightness of ring decreased with the value of GB coupling parameter $\alpha$ increased\footnote{The difference of ring for different value of $\alpha$ maybe be very small, one can see this difference by magnifying those rings.}. At other observed positions, with the increase of $\alpha$ value, the observation brightness of the left area in the image will decrease rapidly, while that of the right area will decrease relatively slowly. For example, when the observation angle is $\theta=60^\circ$, the size and brightness of the bright light arcs in the left area obtained for $\alpha=-0.05$ are significantly stronger than $\alpha=0.05$.
In Fig.4 (c), the observation area on the left side of the image has almost no observable brightness in the case of observation angle is $\theta=60^\circ$.
Not only that, it is interesting that there are two light points for $\alpha=0.1$, where one of which the brightness is very small that it can not be seen as $\theta=90^\circ$.
Combined with above facts, we can conclude that the feature of holographic images can be regarded as an effective tool for revealing the geometric characteristics of black hole.
\begin{figure}[!h]
\makeatletter
\renewcommand{\@thesubfigure}{\hskip\subfiglabelskip}
\makeatother
\centering 
\subfigure[(a): $\alpha = 0.05, r_h=0.7$]{
\setcounter{subfigure}{0}\subfigure[$\theta_{obs}=0^\circ$]{\includegraphics[width=.22\textwidth]{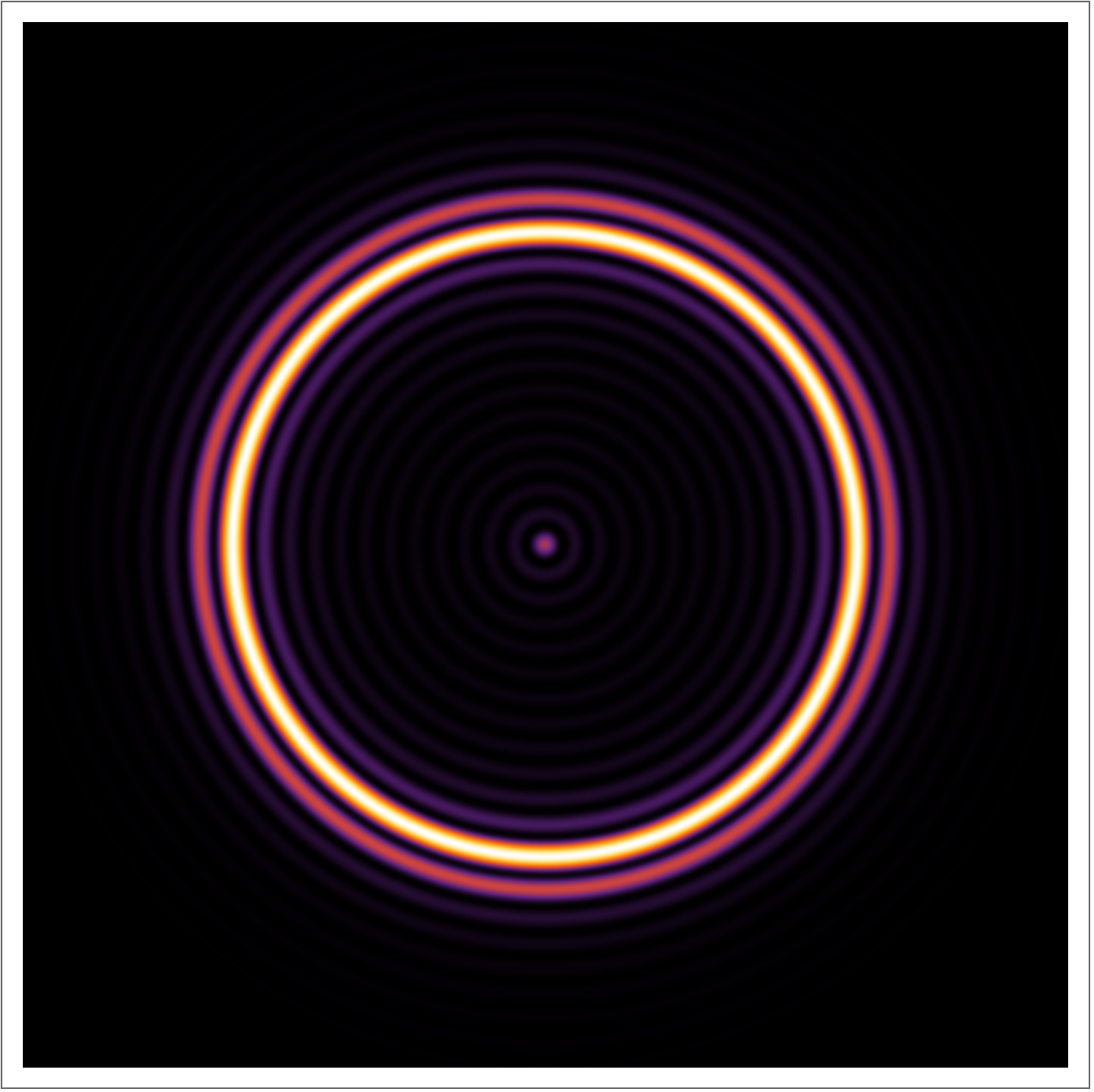}}
\subfigure[$\theta_{obs}=30^\circ$]{\includegraphics[width=.22\textwidth]{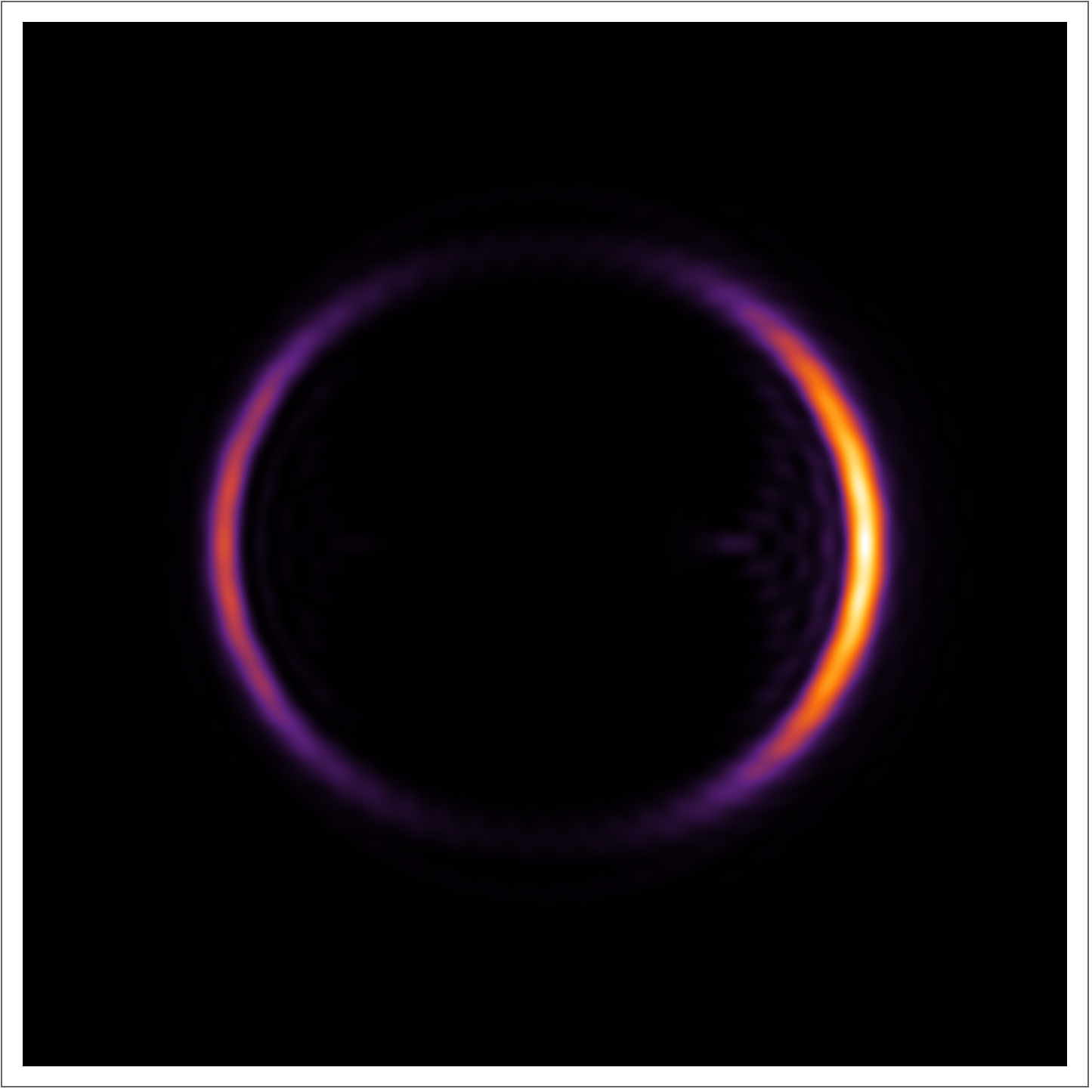}}
\subfigure[$\theta_{obs}=60^\circ$]{\includegraphics[width=.22\textwidth]{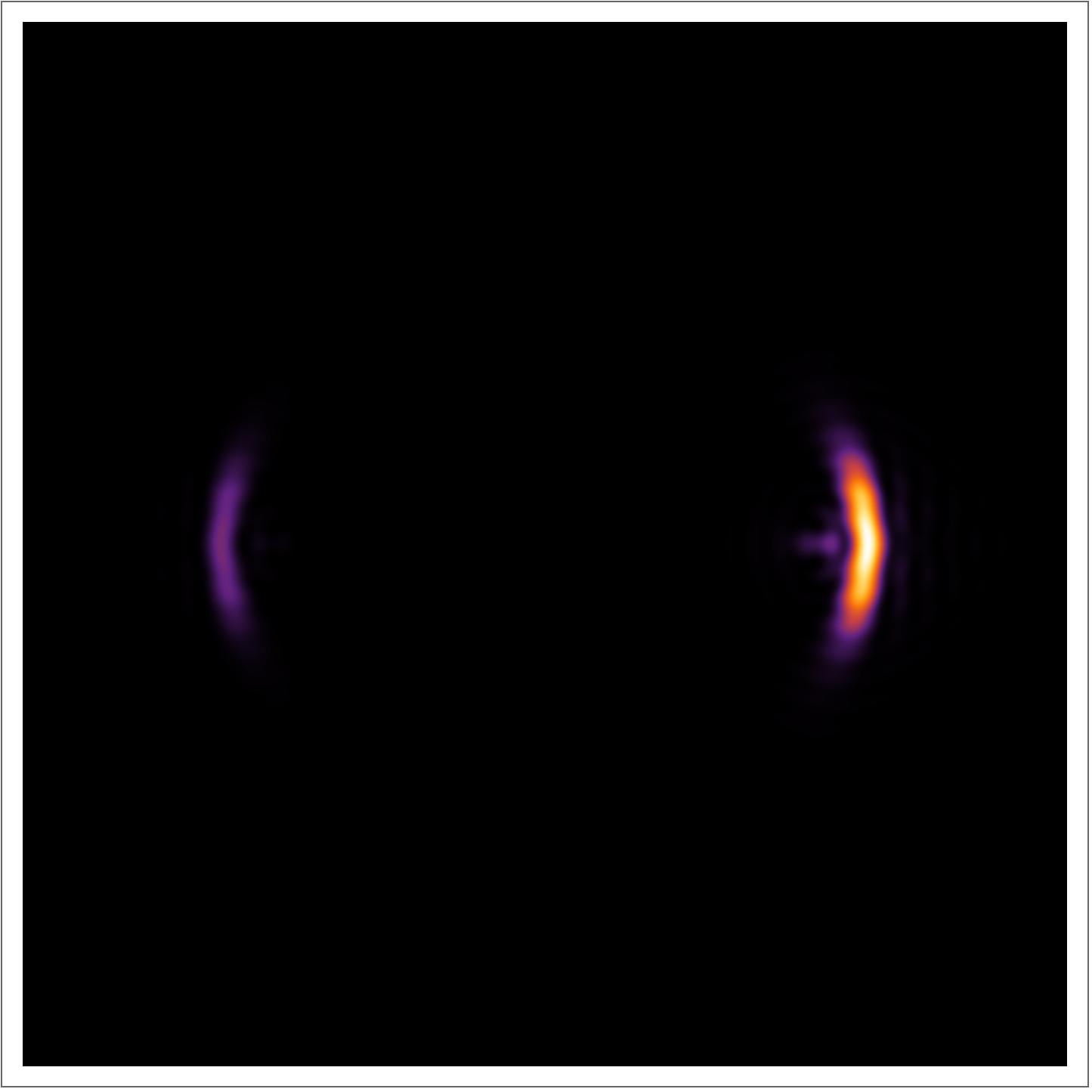}}
\subfigure[$\theta_{obs}=90^\circ$]{\includegraphics[width=.22\textwidth]{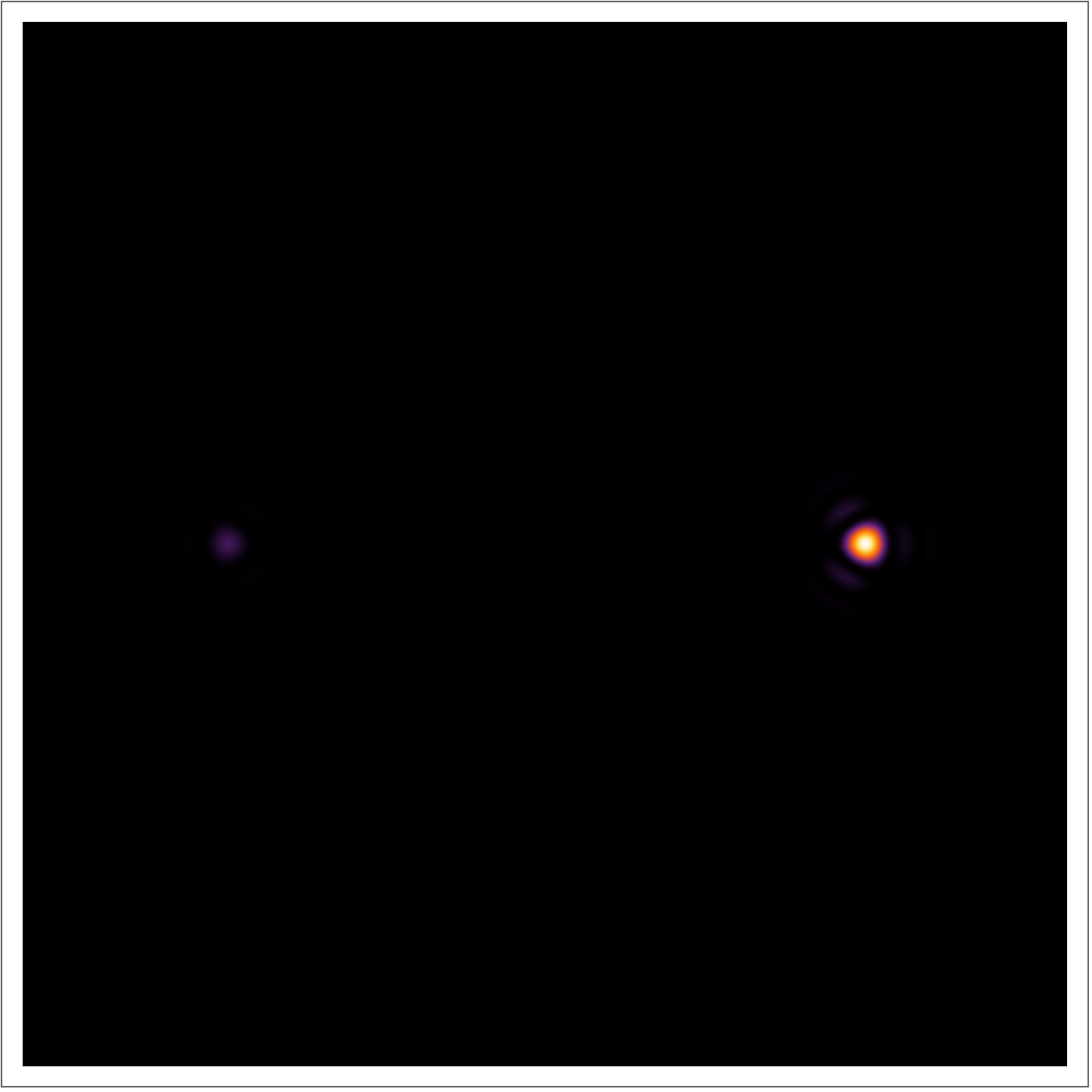}}}
\end{figure}
\begin{figure}[!h]
\makeatletter
\renewcommand{\@thesubfigure}{\hskip\subfiglabelskip}
\makeatother
\centering 
\subfigure[(b): $\alpha = 0.05, \omega=50$]{
\setcounter{subfigure}{0}\subfigure[$\theta_{obs}=0^\circ$]{\includegraphics[width=.22\textwidth]{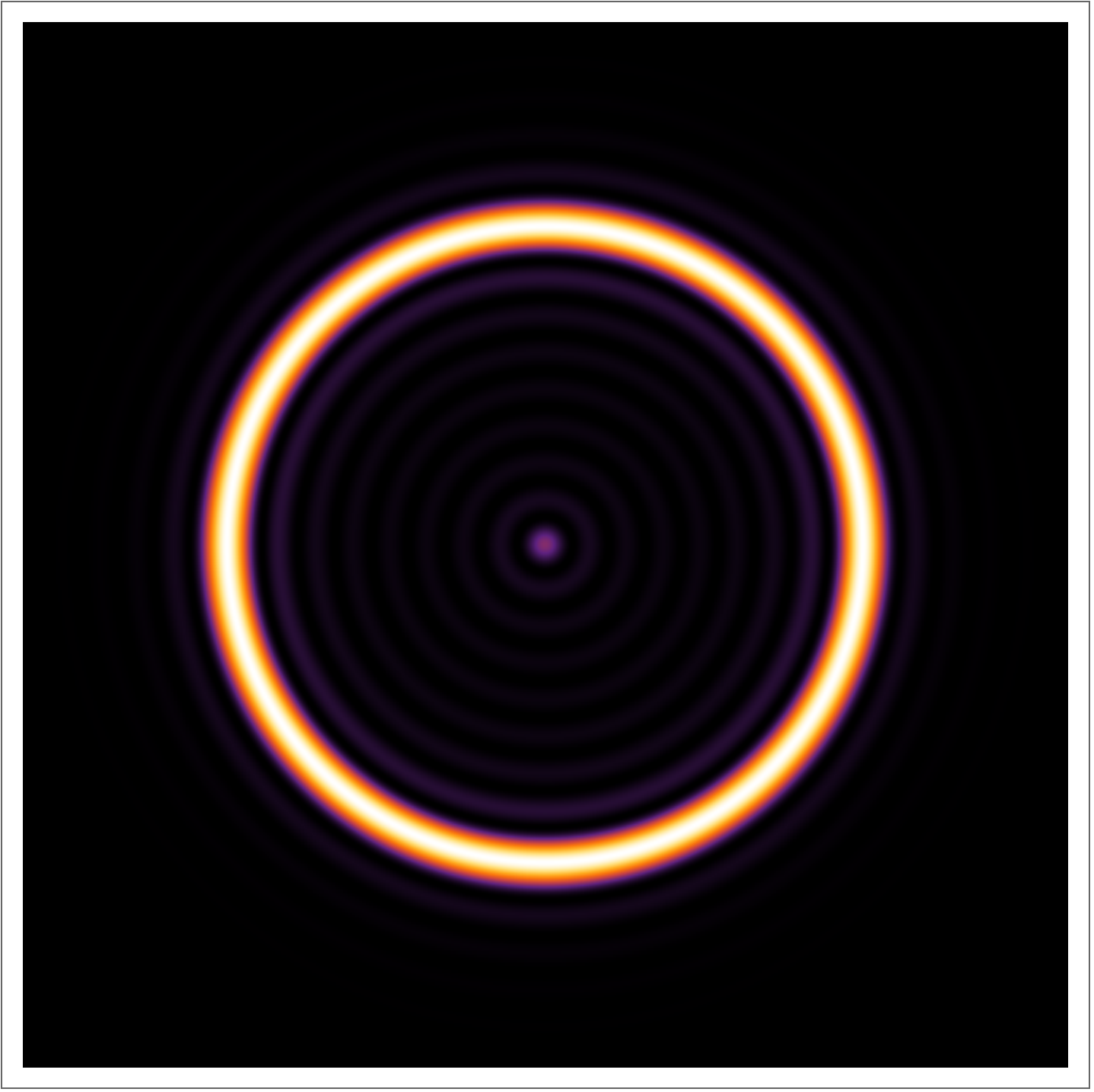}}
\subfigure[$\theta_{obs}=30^\circ$]{\includegraphics[width=.22\textwidth]{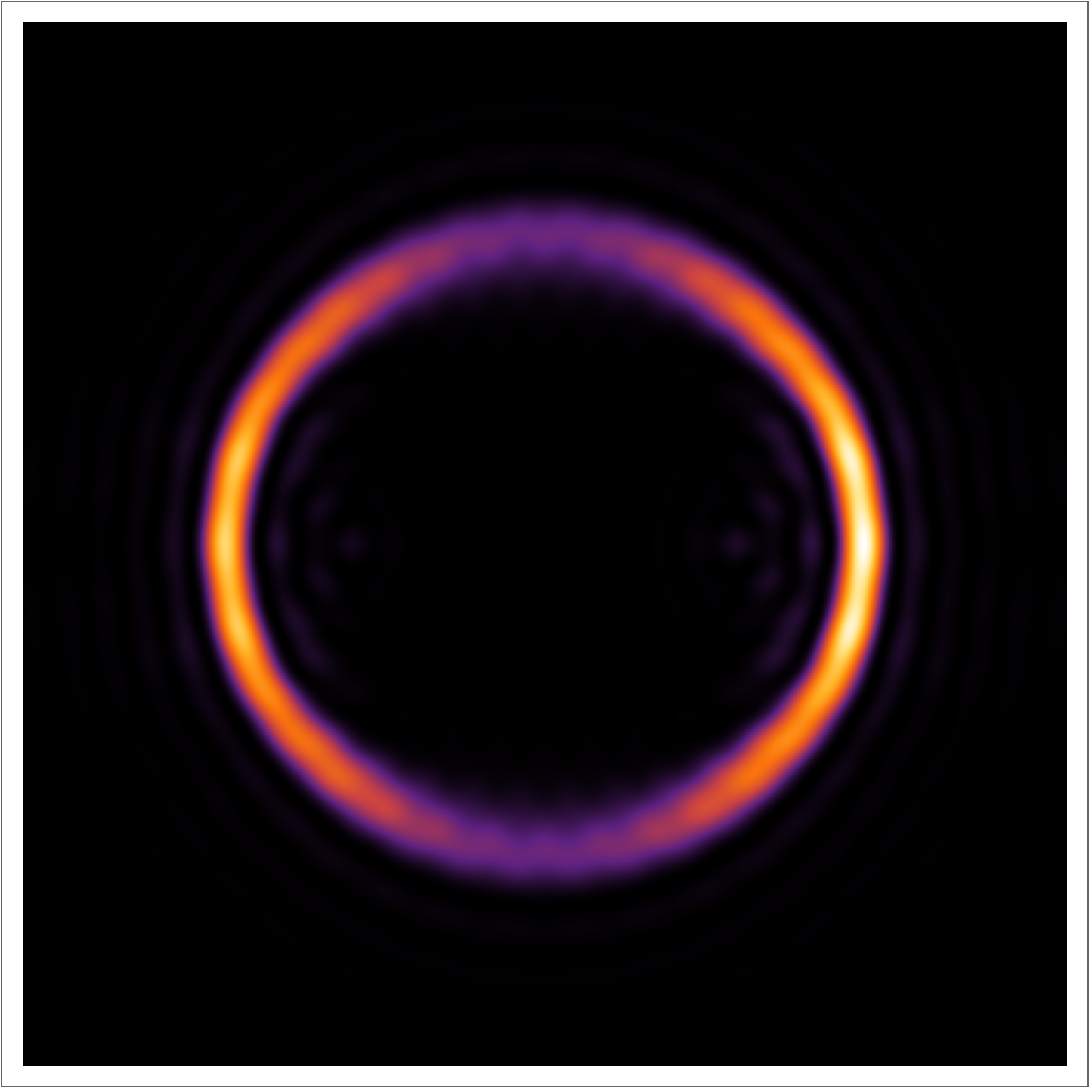}}
\subfigure[$\theta_{obs}=60^\circ$]{\includegraphics[width=.22\textwidth]{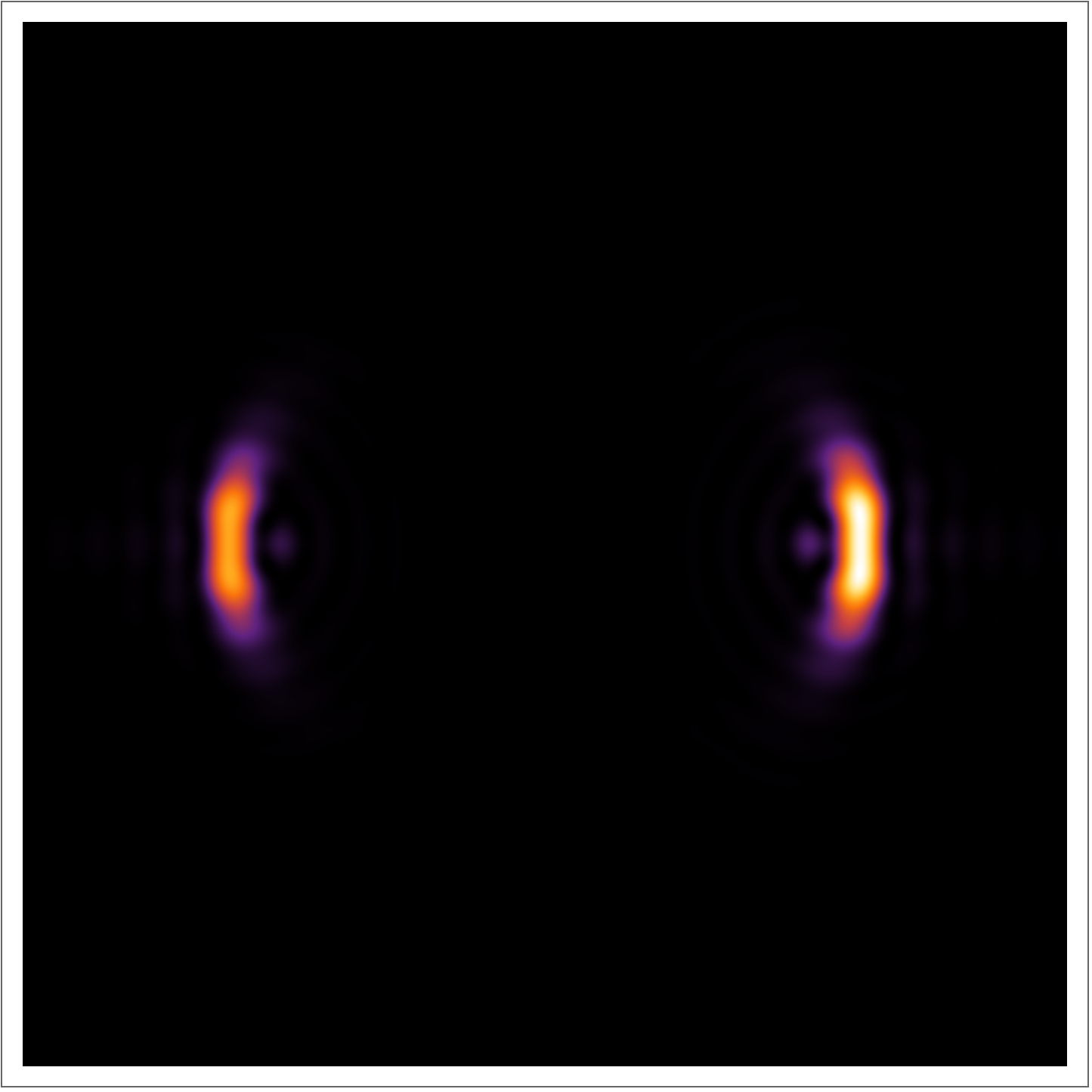}}
\subfigure[$\theta_{obs}=90^\circ$]{\includegraphics[width=.22\textwidth]{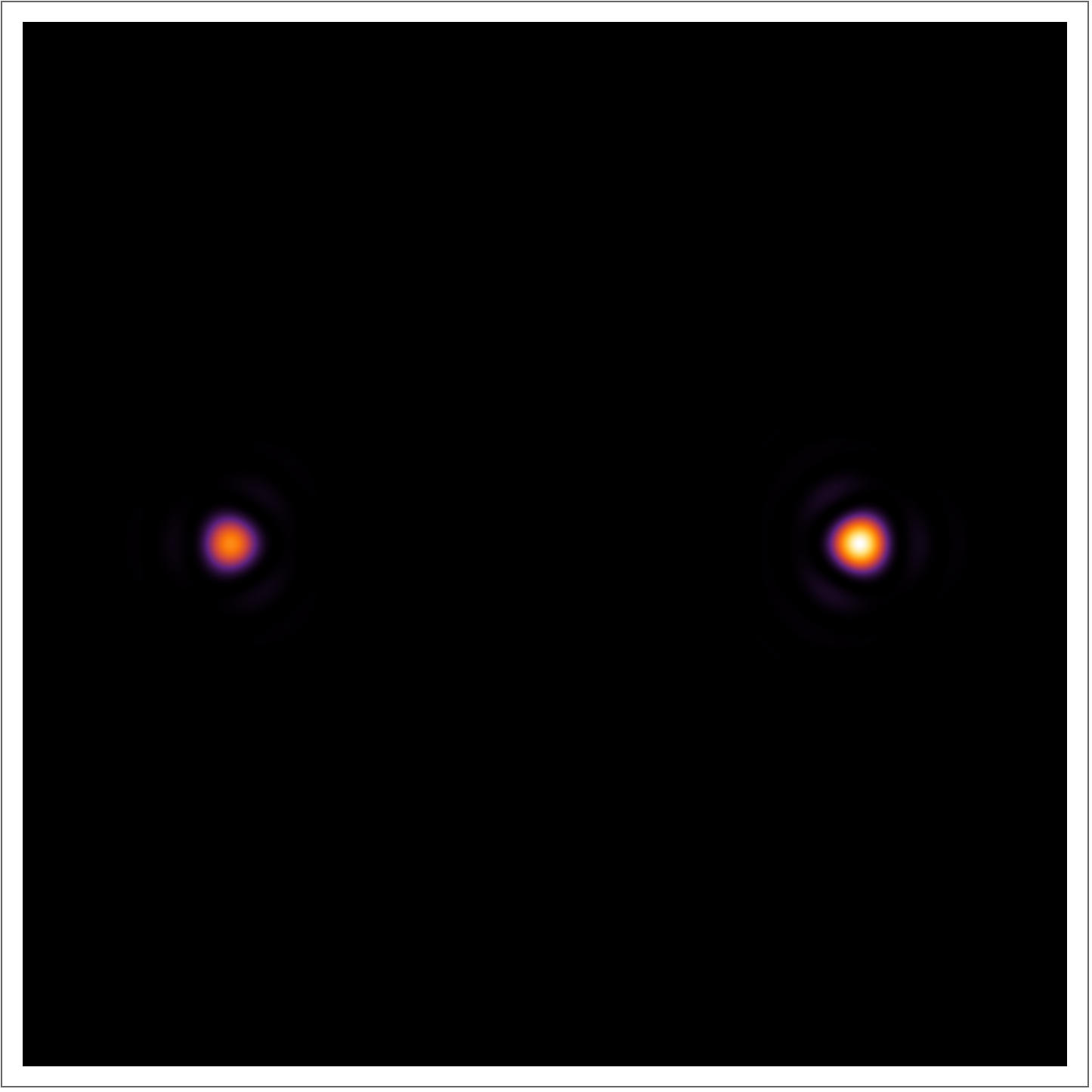}}}
\end{figure}
\begin{figure}[!h]
\makeatletter
\renewcommand{\@thesubfigure}{\hskip\subfiglabelskip}
\makeatother
\centering 
\subfigure[(c): $\alpha = 0.05, d=0.4$]{
\setcounter{subfigure}{0}\subfigure[$\theta_{obs}=0^\circ$]{\includegraphics[width=.22\textwidth]{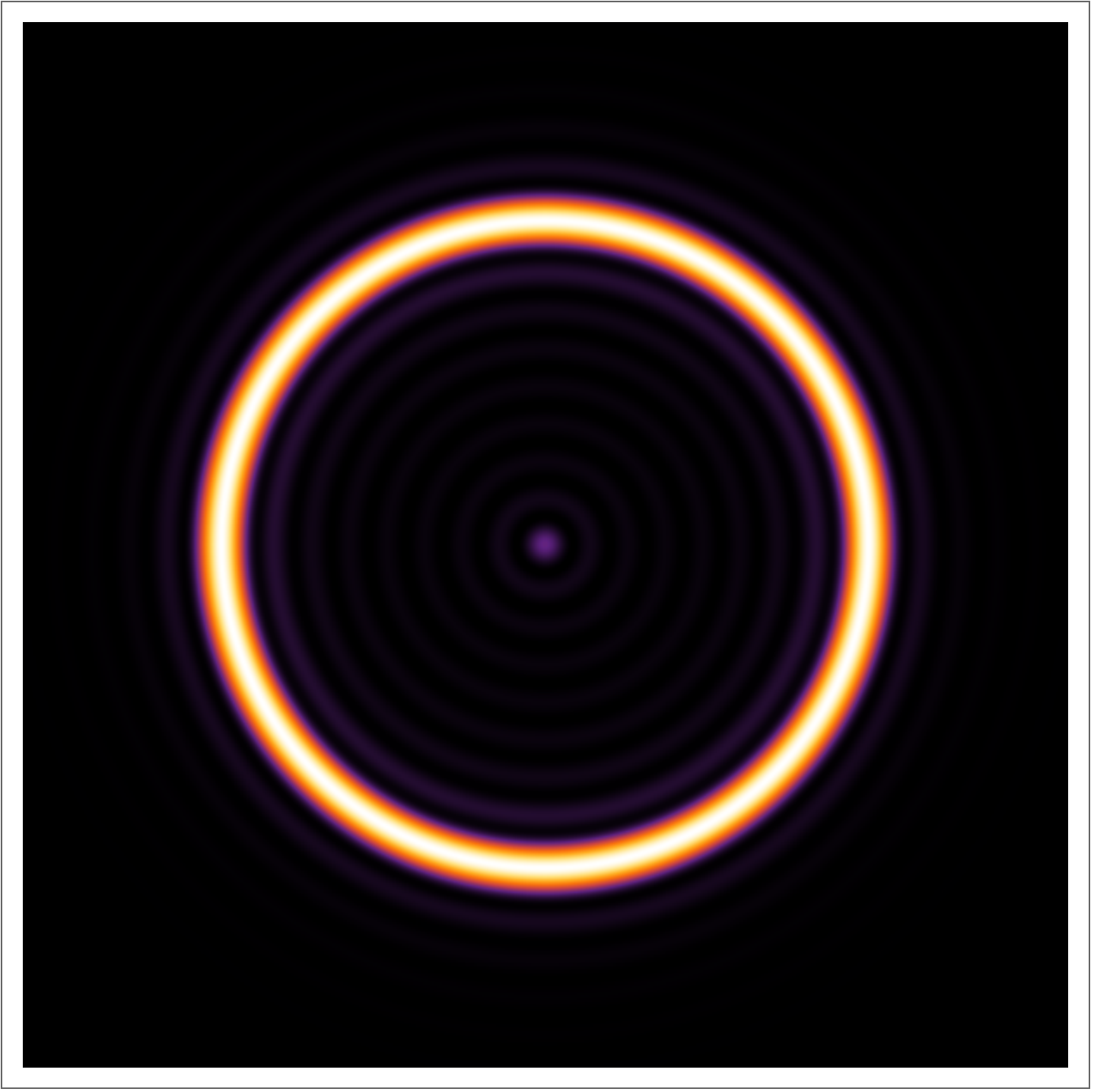}}
\subfigure[$\theta_{obs}=30^\circ$]{\includegraphics[width=.22\textwidth]{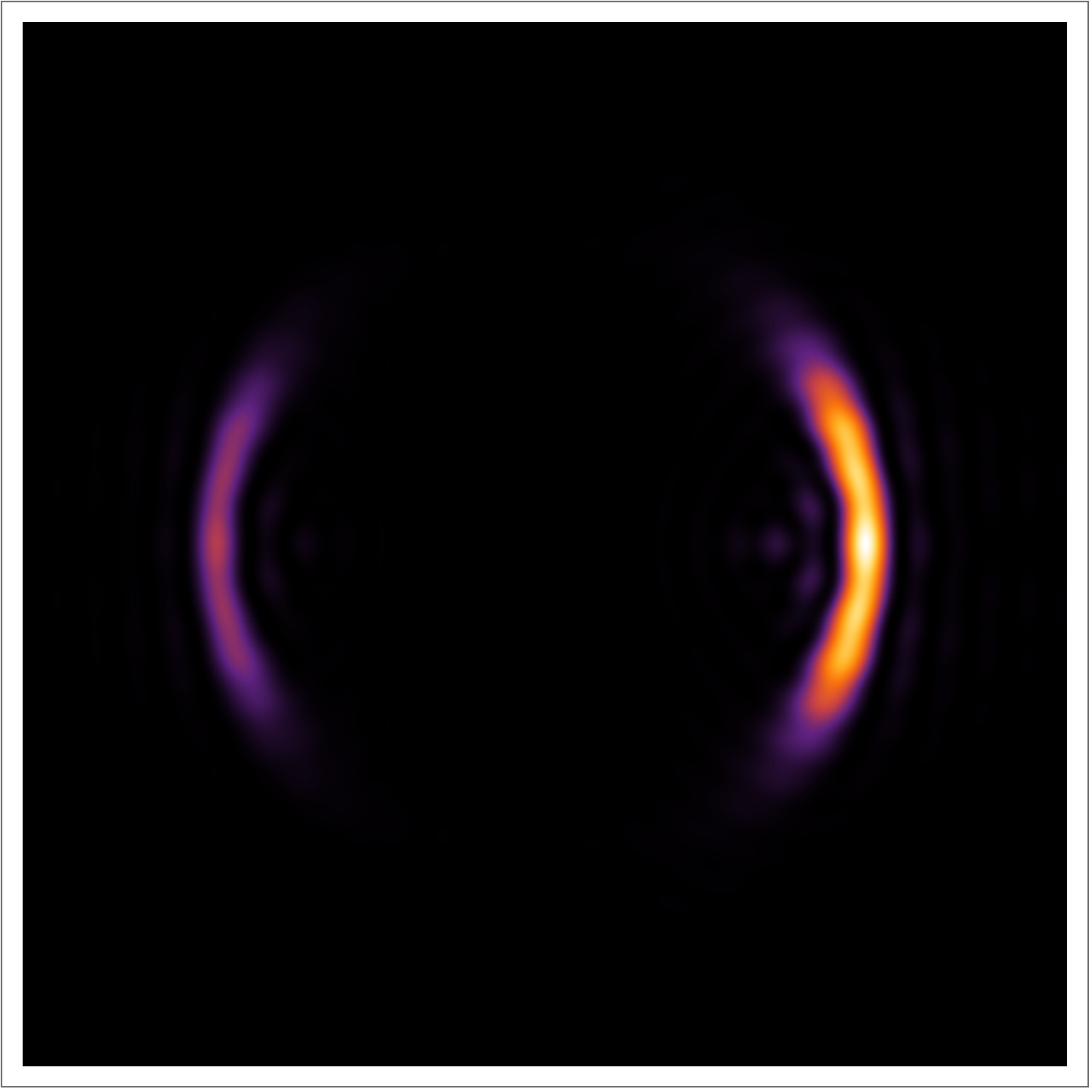}}
\subfigure[$\theta_{obs}=60^\circ$]{\includegraphics[width=.22\textwidth]{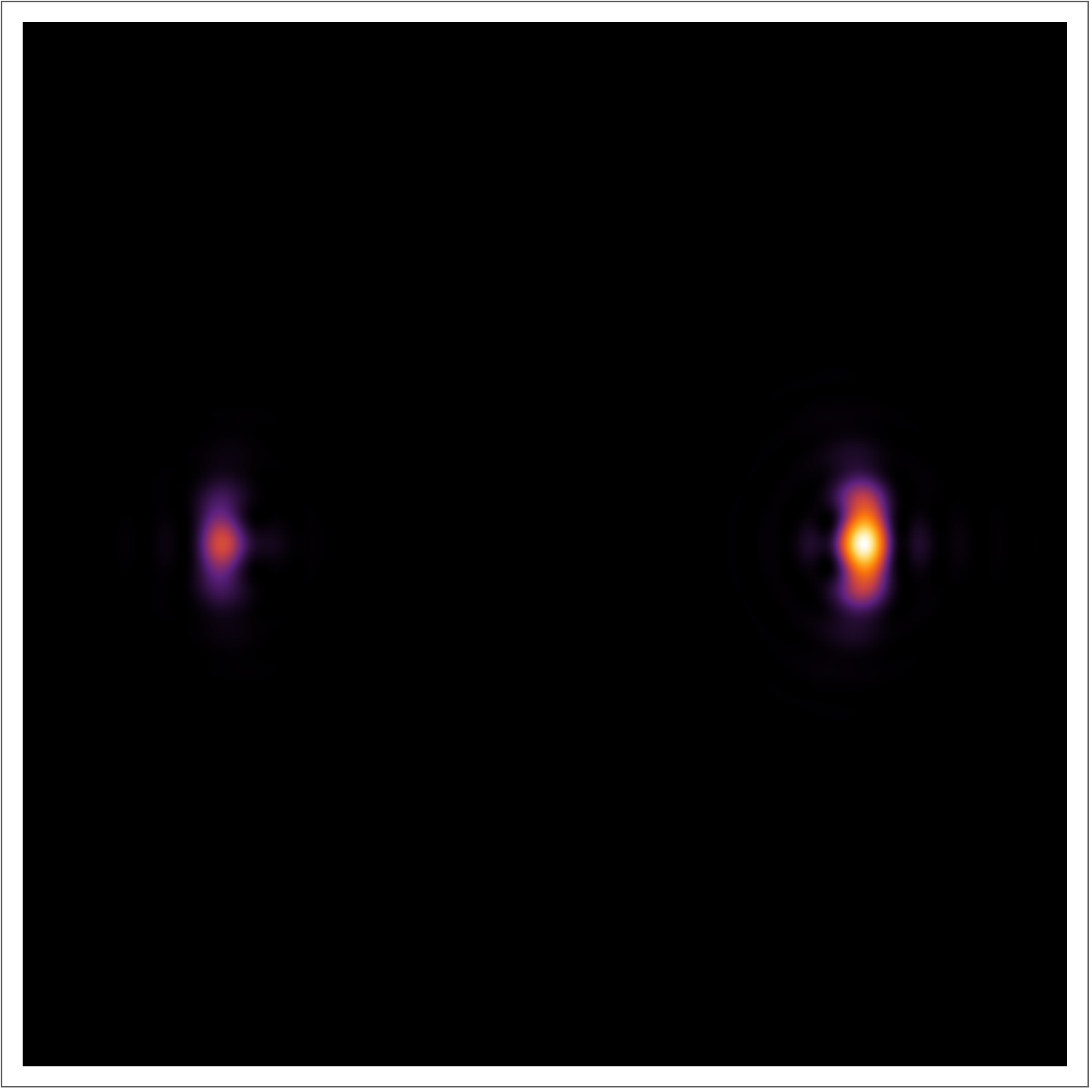}}
\subfigure[$\theta_{obs}=90^\circ$]{\includegraphics[width=.22\textwidth]{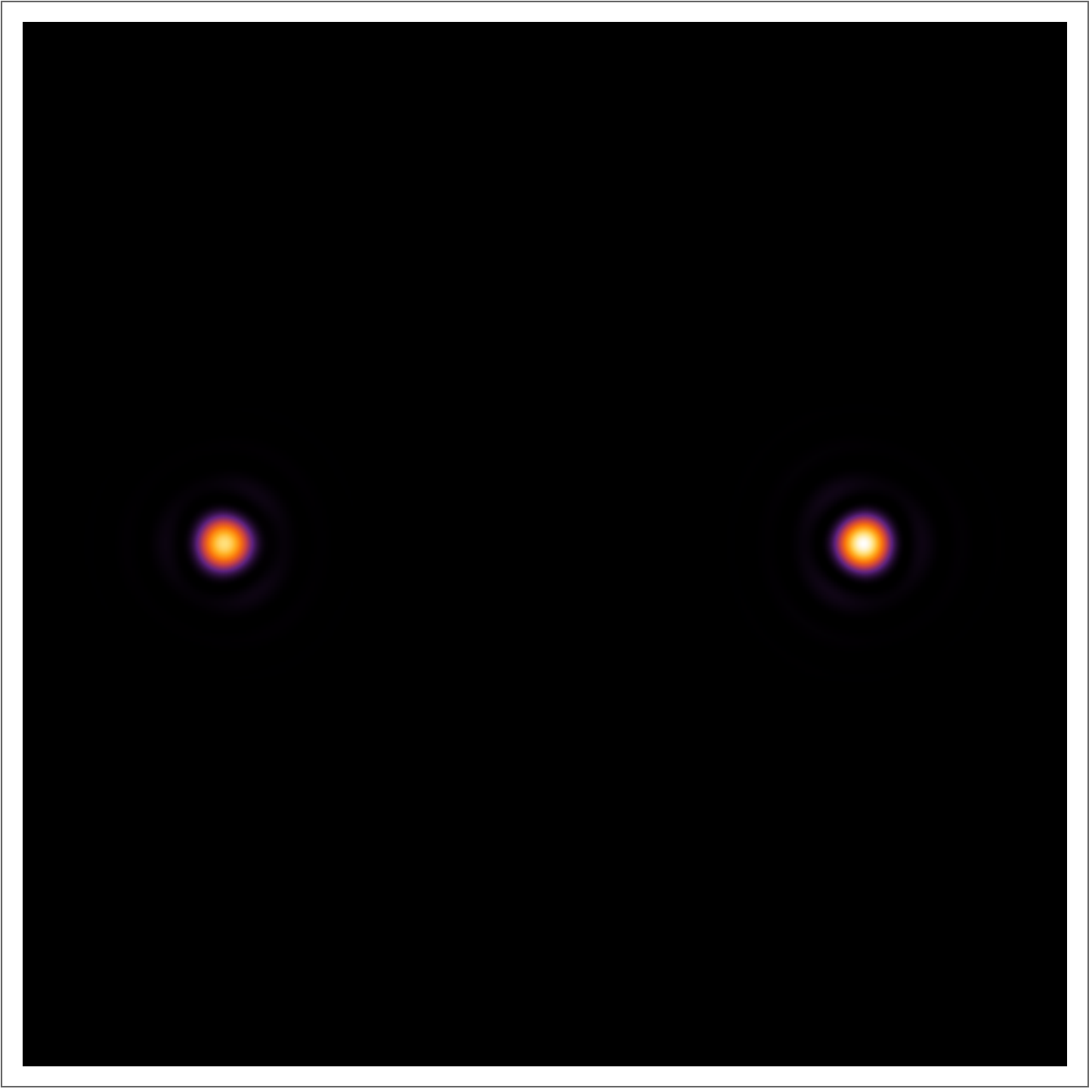}}}
\caption{\label{2fig666} The holographic images for $\alpha = 0.05$, and other corresponding parameters. }
\end{figure}

In addition, we also consider whether the changes of wave source and optical system will affect the characteristics of the holographic Einstein image, and shown in Fig.5.
According to Fig.\ref{2fig666}, it first shows that the interference pattern is obviously enhanced with the increase of the horizon $r_h$, although the size and width of the holographic ring becomes smaller.
And then, the results further show that the width of holographic ring becomes larger for a lower value of frequency of Gussian wave packet source $\omega$ and the lens parameter $d$.
Also, it can be found that the radius of it seems hardly changed for $\omega$, but decreased with the decrease of $d$.
Similar to $r_h$, it is true that the concentric striped patterns as well as holographic images are more indistinct for $\omega$ and $d$.
More importantly, we find from the subfigure (c) of Fig.\ref{2fig666} that the luminosity of ring deformed more quickly in the case $\theta_{obs}=30^\circ$ by comparing with the subfigure (a) of Fig.4. In addition, at the position $\theta_{obs}=90^\circ$, one may also observed two light points in the subfigures (a) and (b) of Fig.\ref{2fig666}, rather than only one light point which shown in the subfigure (c) of Fig.\ref{1fig666}.
In view of this, one can see that the holographic images of AdS black hole can not only characterize the geometric of black hole, but also closely related to the properties of lens and wave packet source.

For the brightest ring in the image, it corresponds to the position of the photon sphere of the black hole. We will verify this bright ring in the image from the perspective of optical geometry. Therefore, we will focus on the light deflection caused by the four-dimensional  AdS Gauss-Bonnet black hole, that is, and the crux is the motion  of photons around the black hole. It worth note that, the Hamilton-Jacobi equation of  null geodesic in general curved spacetime  can be given by the massless Klein-Gordon equation through the the eikonal approximation. In this system, the Lagrangian $\mathcal {L}$ of the photon is
\begin{align}
2 \mathcal{L} = g_{\mu\nu} \dot{x}^{\mu } \dot{x}^{\nu },\label{Lag}
\end{align}
in which the term of $\dot{x}^{\mu }$ is the four-velocity of the photon.  By considering the general spherically symmetric spacetime, we can  restrict our discussion to the equatorial plane, namely, $\theta=\frac{1}{2} \pi$.
For the metric coefficients in Eq.(\ref{eq3}), it is not rely on the time $t$ and azimuthal angle $\phi$. Hence, one can  obtain two constants of motion  $\hat{e}$ and $\ell$ which are related to the energy and angular momentum of a given geodesic, which is
\begin{align}\label{Eq23}
\hat{e}=F (r) \dot{t} = constant, \quad
\ell=r^2 \dot {\phi} = constant.
\end{align}
Based on the null geodesic $g_{\mu\nu} \dot{x}^{\mu } \dot{x}^{\nu }=0$. And,  by introducing the affine parameter $\lambda=\lambda/\ell$, we further have
\begin{align}\label{Eq24}
\dot{r}^2=\frac{1}{b^2}-V_{eff}.
\end{align}
In the above equation, $b=\frac{\ell}{\hat{e}}$ is called the impact parameter. The behavior of the  geodesic lines depends on the choice of the impact parameter $b$. In addition, the term of $V_{eff}$ is the effective potential, which is express as
\begin{align}
V_{eff}=\frac{F(r)}{r^2}. \label{V2}
\end{align}
We plot the effective potential $V_{eff}$ as a function of radius $r$ for the value of $M = 1$, in Fig. 6.
The photon sphere orbit conditions are $\dot{r}=0$ and $\ddot{r}=0$, which mean that the effective potential is
\begin{align}
V_{eff}(r_{ph})=\frac{1}{b_{ph}^2},\quad
V'_{eff}(r_{ph})=0, \label{V3}
\end{align}
Therefore, the  position of the maximum effective potential $V_{max}$ is the position of the photon sphere $r_{ph}$, and the impact parameter is
\begin{align}\label{B1}
b_{ph}=\frac{1}{\sqrt{V_m}}
\end{align}.

\begin{figure}[!h]
\makeatletter
\renewcommand{\@thesubfigure}{\hskip\subfiglabelskip}
\makeatother
\centering 
\subfigure[]{
\setcounter{subfigure}{0}\subfigure[]{\includegraphics[width=0.4\textwidth]{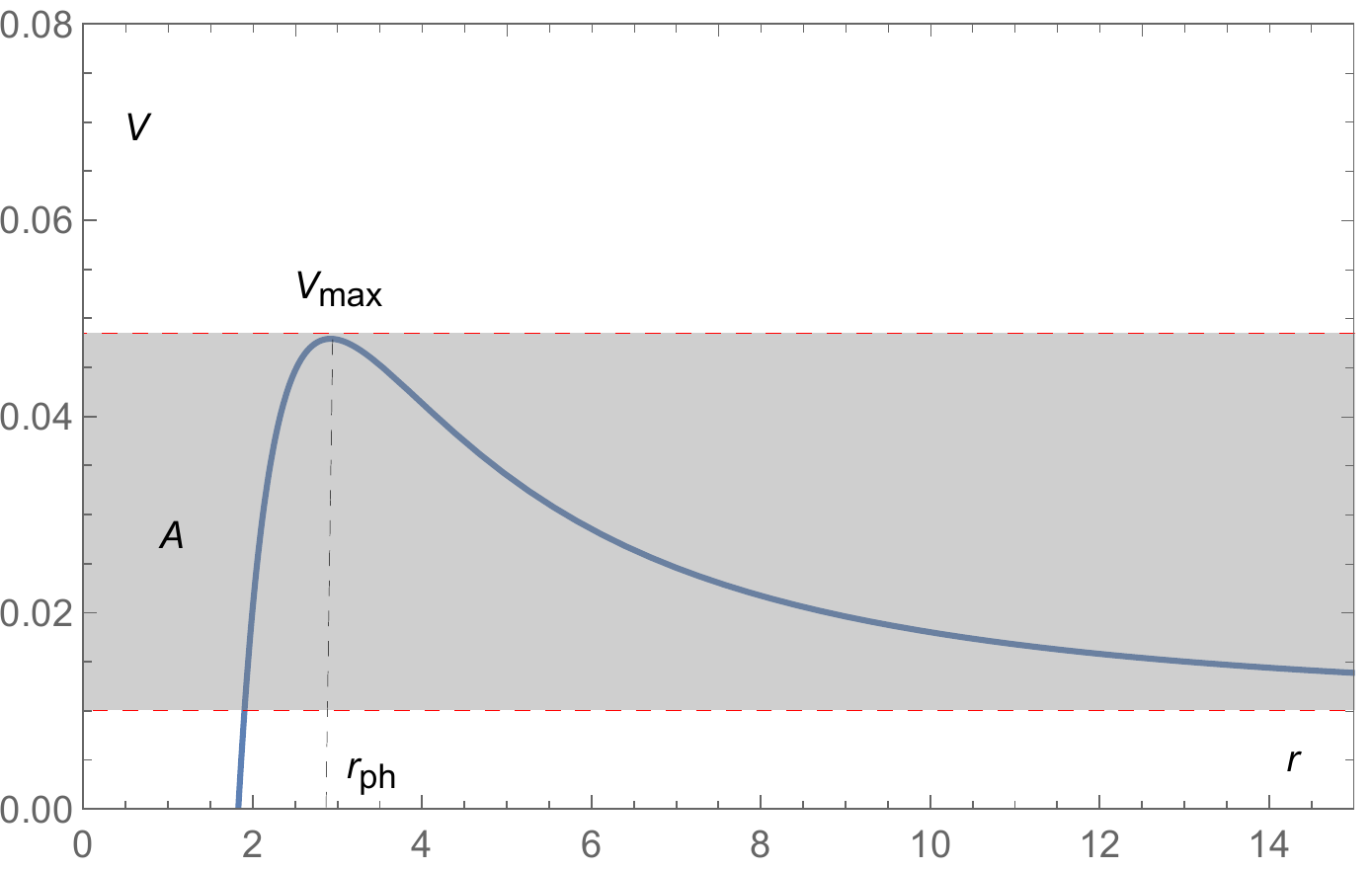}}}
\caption{\label{figres111} The effective potential for $l=10$. }
\end{figure}

If the observer on one side of the AdS boundary wants to capture the photons emitted from the other side of the boundary, the impact parameter need to meet condition $b>b_{ph}$. Otherwise, the photon will fall directly into the black hole in the case of $b<b_{ph}$. Because the limiting factor $ \lim_{r\rightarrow \infty}V(r) =1$\footnote{For convenience, we have employ $l=10$ to show the effective potential more obviously in Fig.6. And here, we have used the $l=1$  in the text}, the photons that can be captured by the observer belong to the A region, i.e., $1<\frac{1}{b^2}<V_{max} $, which is shown in Fig.6. In particular, at the position of the photon sphere $b=b_{ph}$, photons will neither escape from the black hole nor fall into the black hole, and are in a state of constant rotation around the black hole. Hence, the closer the value of impact parameter is to $b_{ph}$, the more cycles of the photon rotates around the black hole. In Fig.7, we show the schematic diagram of photons starting from the south Pole, rotating around the black hole for one time and arriving at the north pole.
\begin{figure}[!h]
\makeatletter
\renewcommand{\@thesubfigure}{\hskip\subfiglabelskip}
\makeatother
\centering 
\subfigure[]{
\setcounter{subfigure}{0}\subfigure[(a)]{\includegraphics[width=0.42\textwidth]{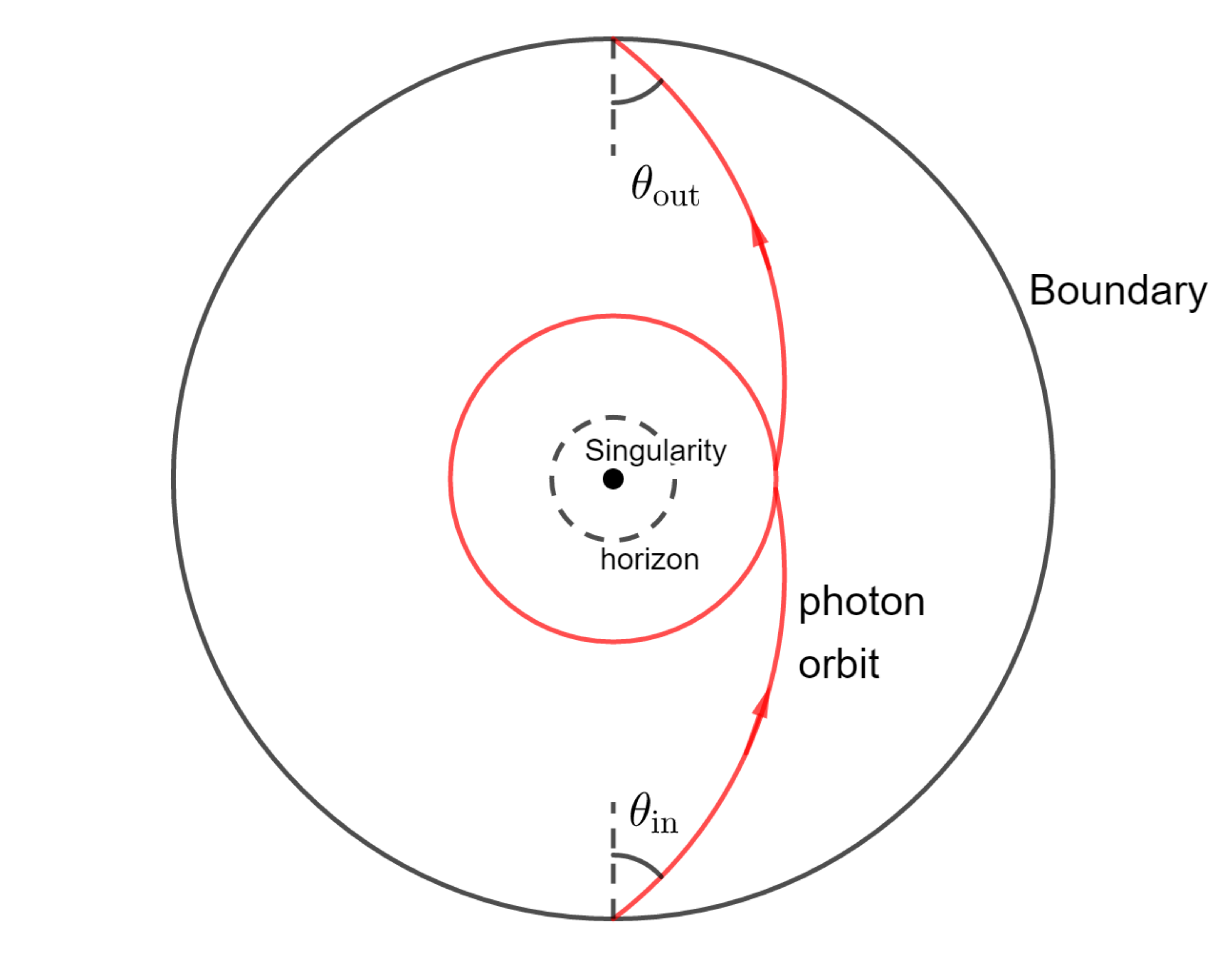}}
\subfigure[(b)]{\includegraphics[width=0.48\textwidth]{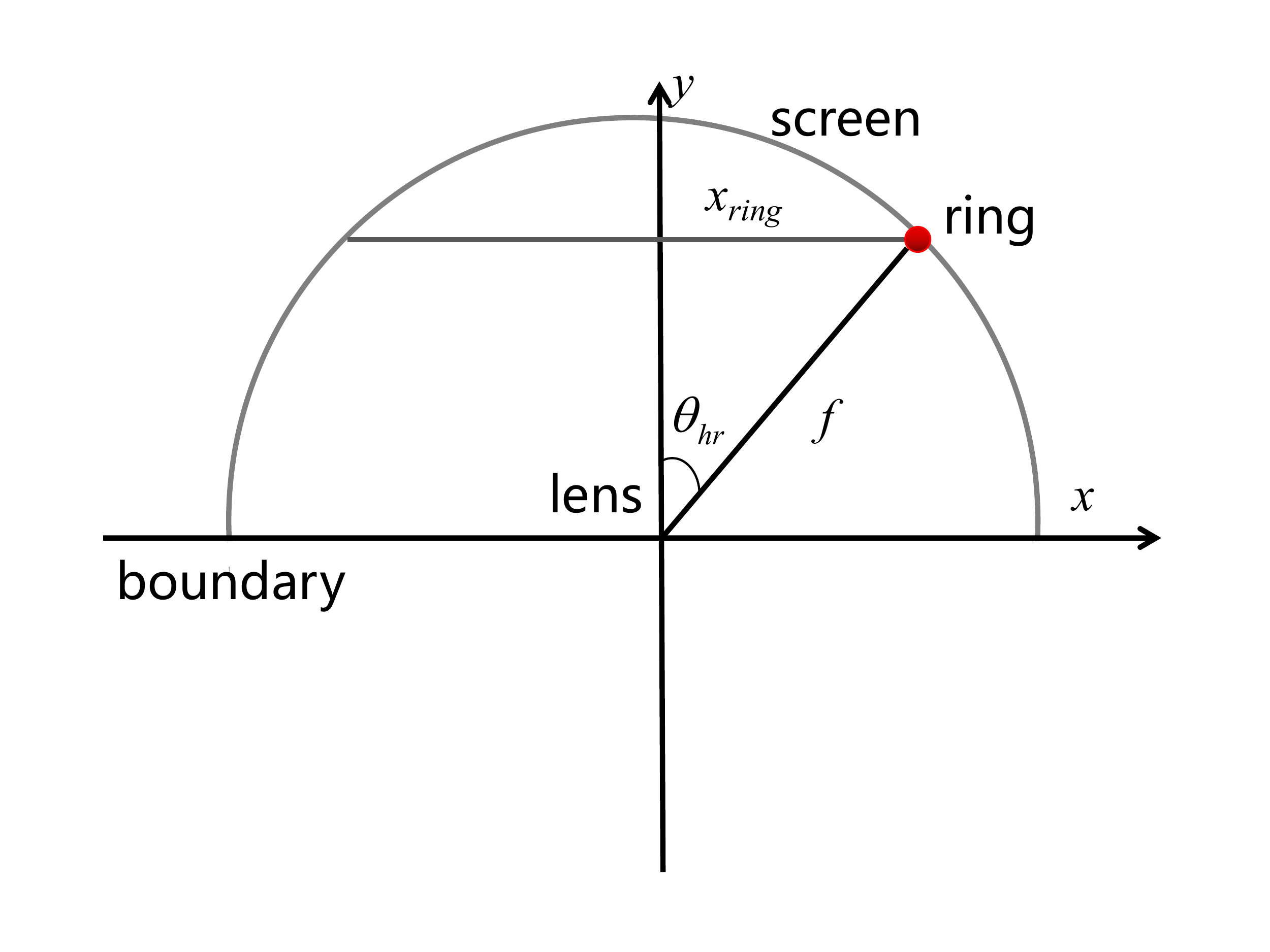}}}
\caption{\label{figures11} (a) Schematic diagram of the orbit of the incident photon rotating once around the black hole; (b) the relation between $\theta_{hr}$ and $x_{ring}$. }
\end{figure}

It is evident that the incident angle  between the photon orbit and radial direction is equal to the emitted angle $\theta_{in}=\theta_{out}$,  and one can get \cite{Glavan:2019inb}
\begin{align}
\sin\theta_{in}=\frac{\ell}{\mathit{\hat{e}}}.\label{S1}
\end{align}

Therefore,due to the axisymmetry, it is obvious that the observer can clearly see a bright ring with its radius is closely related to $r_{ph}$ the geometrical optics. Meanwhile, we can also defined an angle to characterize the radius of holographic ring obtained in Fig.\ref{figures11}(b), which reads
\begin{align}\label{eq28}
\sin \theta_{\rm hr} = \frac{x_{ring}}{f},
\end{align}
By carrying out the similar approach in \cite{Glavan:2019inb}, one can also employ the spherical harmonics $Y_{lm}(\theta,\varphi)$ in Eq.(\ref{eq21}) to find the position of peak of the image, which is $\frac{\ell}{\mathit{\hat{e}}}=\frac{x_{ring}}{f}$. So, this means that the position of photon ring obtained from the geometrical optics is full inconsistence with that of the holographic ring.

\begin{figure}[!h]
\makeatletter
\renewcommand{\@thesubfigure}{\hskip\subfiglabelskip}
\makeatother
\centering 
\subfigure[]{
\setcounter{subfigure}{0}\subfigure[]{\includegraphics[width=0.5\textwidth]{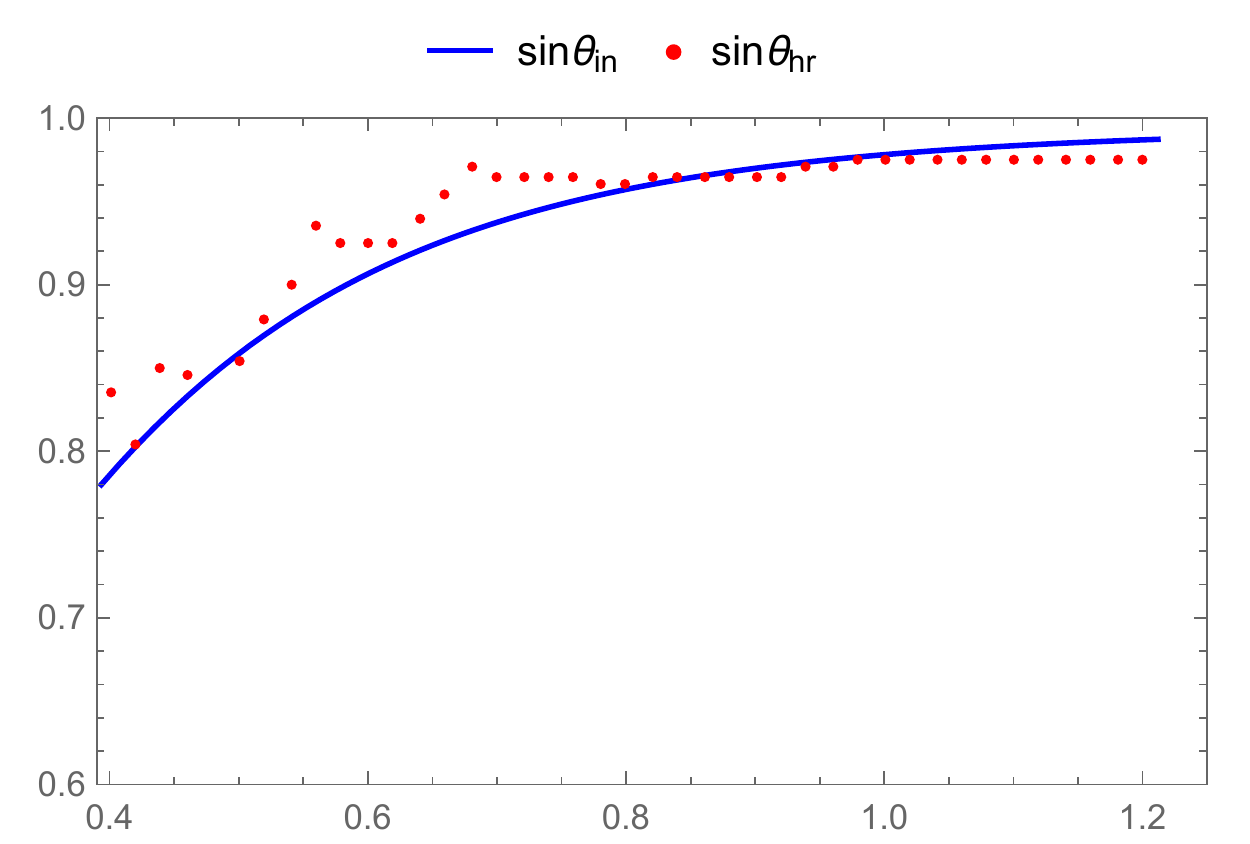}}}
\caption{\label{figres111} Comparison of ring angle $\theta_{hr}$ and angle of photon sphere $\theta_{in}$ in Gauss-Bonnet AdS black hole, where $\omega = 75, r_h=0.6, M=1$. }
\end{figure}

To illustrate relationship $\sin\theta_{in}=\sin\theta_{hr}$, we take the case $\alpha=0.01$ as an example to numerically present the ring angle $\theta_{in}$ and angle of photon sphere $\theta_{hr}$ in Gauss-Bonnet AdS black hole, where the red dot represents the radius of holographic ring $\theta_{hr}$, the blue line is the location of photon sphere. From Fig.(\ref{figres111}), it is obvious that the red dot always located on the blue line or its vicinity, this confirm fact that the photon ring's position given by the geometrical optics is always coincide with that of the holographic ring.

\section{Conclusions and discussions}
\label{Concl}
In the framework of Gauss-Bonnet gravity, we have studied the holography images of a  black hole in the bulk from a given response function on AdS boundary. By considering the oscillating Gaussian source produced at a point on the boundary, we have first used the AdS/CFT dictionary to compute the response function,  which is  shown in the  another side of the AdS boundary.
The result show that there always exists the diffraction pattern of total response function after the scalar wave passed through the black hole.
And, the absolute amplitude of it does not only closely depend on the spacetime geometry, i.e., the GB coupling parameter $\alpha$ of black hole, but the properties of the source, i.e., the frequency $\omega$ of the wave. In particular, the increase of the parameters $\alpha$, $\omega$ and $r_h$ all decreased the strength of the absolute amplitude of total response function.

Although we have obtained the diffraction pattern of the response function, it does not directly reflect the black hole information.
After performing the Fourier transformation to the response function, the Einstein images of AdS black hole can be observed with an optical system consisting of lens and screen.
It turns out that when the observer located at the north pole($\theta_{obs}=0^\circ$) and the source is located at the south pole, the image of AdS black hole appears as a bright ring, which is the holographic Einstein ring.
This ring surrounded with a series of the concentric stripes, which correspond to the diffraction pattern of total response function. At the center of ring, there is a bright spot which called as the Poisson-like spot that caused by the diffraction of scalar wave. By comparing the obtained images of AdS black hole, one can find that the size and width of the ring are  closely related to the parameters of black hole, optical system and wave source.
For example, the size of holographic Einstein ring decreases with the increase of GB coupling parameter $\alpha$ and $d$, but hardly changed for $\omega$.
For the width, it seems hardly influenced with the GB coupling parameter  $\alpha$, but will becomes larger for a lower value of $\omega$ and $d$.
When the observer is not in the north pole, we can see that the bright ring in the image will gradually evolve into two bright light arcs and finally form two light spots with the increase of observation angle.
Similarly, under different observation angles, one will also observe different holographic images of black hole for different values of parameters of black hole, optical system and wave source.
Finally, we can conclude that the holographic images can be used as an effective tool to distinguish different types of black holes for the fixed wave source and optical system.
Obviously, it is also very interesting to further study the holographic images in other modified gravity theories.
And, it should be noted that that the holographic rings has been extended to case of Maxwell field\cite{Kaku:2021xqp}.
So, it is also interesting to extend this method to the case of other different fields.

\vspace{10pt}

\noindent {\bf Acknowledgments}

\noindent
This work is supported by the National Natural Science Foundation of China (Grant Nos. 11875095 and 11903025), and by the starting fund of China West Normal University (Grant No.18Q062), and by the Sichuan Youth Science and Technology Innovation Research Team (21CXTD0038), and by the Chongqing science and Technology Bureau (cstc2022ycjh-bgzxm0161), and by the Natural Science Foundation of SiChuan Province(2022NSFSC1833), and by Sichuan Science and Technology Program( 2023NSFSC1352).




\begin{thebibliography}{99}

\bibitem{maldacena1999AdSCFT}
J.~M.~Maldacena,
``The Large N limit of superconformal field theories and supergravity,''
Adv. Theor. Math. Phys. \textbf{2}, 231-252 (1998).


\bibitem{aharony2000large}
O.~Aharony, S.~S.~Gubser, J.~M.~Maldacena, H.~Ooguri and Y.~Oz,
``Large N field theories, string theory and gravity,''
Phys. Rept. \textbf{323}, 183-386 (2000).


\bibitem{natsuume2015ads}
M.~Natsuume,
``AdS/CFT Duality User Guide,''
Lect. Notes Phys. \textbf{903}, pp.1-294 (2015).


\bibitem{erlich2005QCD}
J.~Erlich, E.~Katz, D.~T.~Son and M.~A.~Stephanov,
``QCD and a holographic model of hadrons,''
Phys. Rev. Lett. \textbf{95}, 261602 (2005).


\bibitem{hartnoll2009CMT}
S.~A.~Hartnoll,
``Lectures on holographic methods for condensed matter physics,''
Class. Quant. Grav. \textbf{26}, 224002 (2009).

\bibitem{gubser2008breaking}
S.~S.~Gubser,
``Breaking an Abelian gauge symmetry near a black hole horizon,''
Phys. Rev. D \textbf{78}, 065034 (2008).

\bibitem{hartnoll2008HS}
S.~A.~Hartnoll, C.~P.~Herzog and G.~T.~Horowitz,
``Holographic Superconductors,''
JHEP \textbf{12}, 015 (2008).

\bibitem{hartnoll2008building}
S.~A.~Hartnoll, C.~P.~Herzog and G.~T.~Horowitz,
``Building a Holographic Superconductor,''
Phys. Rev. Lett. \textbf{101}, 031601 (2008).

\bibitem{herzog2009holographic}
C.~P.~Herzog, P.~K.~Kovtun and D.~T.~Son,
``Holographic model of superfluidity,''
Phys. Rev. D \textbf{79}, 066002 (2009).

\bibitem{strominger2001ds}
A.~Strominger,
``The dS / CFT correspondence,''
JHEP \textbf{10}, 034 (2001).

\bibitem{bredberg2011lectures}
I.~Bredberg, C.~Keeler, V.~Lysov and A.~Strominger,
``Cargese Lectures on the Kerr/CFT Correspondence,''
Nucl. Phys. B Proc. Suppl. \textbf{216}, 194-210 (2011).



\bibitem{Huang:2004ai}
Q.~G.~Huang and M.~Li,
``The Holographic dark energy in a non-flat universe,''
JCAP \textbf{08}, 013 (2004).

\bibitem{Li:2009zs}
M.~Li, X.~D.~Li, S.~Wang, Y.~Wang and X.~Zhang,
``Probing interaction and spatial curvature in the holographic dark energy model,''
JCAP \textbf{12}, 014 (2009).

\bibitem{Sheykhi:2009zv}
A.~Sheykhi,
``Thermodynamics of interacting holographic dark energy with apparent horizon as an IR cutoff,''
Class. Quant. Grav. \textbf{27}, 025007 (2010).

\bibitem{Micheletti:2009jy}
S.~M.~R.~Micheletti,
``Observational constraints on holographic tachyonic dark energy in interaction with dark matter,''
JCAP \textbf{05}, 009 (2010).

\bibitem{Setare:2010wt}
M.~R.~Setare and M.~Jamil,
``Holographic dark energy with varying gravitational constant in Horava-Lifshitz cosmology,''
JCAP \textbf{02}, 010 (2010).

\bibitem{Huang:2010zzt}
P.~Huang and Y.~C.~Huang,
``A holographic energy model,''
Eur. Phys. J. C \textbf{69}, 503-507 (2010).


\bibitem{Lu:2009iv}
R.~G.~Cai, L.~Li, L.~F.~Li and R.~Q.~Yang,
``Introduction to Holographic Superconductor Models,''
Sci. China Phys. Mech. Astron. \textbf{58}, no.6, 060401 (2015)

\bibitem{Bai:2014poa}
X.~Bai, B.~H.~Lee, M.~Park and K.~Sunly,
``Dynamical Condensation in a Holographic Superconductor Model with Anisotropy,''
JHEP \textbf{09}, 054 (2014).

\bibitem{Aprile:2012sr}
F.~Aprile,
``Holographic Superconductors in a Cohesive Phase,''
JHEP \textbf{10}, 009 (2012).

\bibitem{Cai:2017ihd}
R.~G.~Cai, X.~X.~Zeng and H.~Q.~Zhang,
``Influence of inhomogeneities on holographic mutual information and butterfly effect,''
JHEP \textbf{07}, 082 (2017).


\bibitem{Kusuki:2019zsp}
Y.~Kusuki, J.~Kudler-Flam and S.~Ryu,
``Derivation of holographic negativity in AdS$_3$/CFT$_2$,''
Phys. Rev. Lett. \textbf{123}, no.13, 131603 (2019).

\bibitem{Akers:2019nfi}
C.~Akers, N.~Engelhardt and D.~Harlow,
``Simple holographic models of black hole evaporation,''
JHEP \textbf{08}, 032 (2020).

\bibitem{Bhattacharya:2021jrn}
A.~Bhattacharya, A.~Bhattacharyya, P.~Nandy and A.~K.~Patra,
``Islands and complexity of eternal black hole and radiation subsystems for a doubly holographic model,''
JHEP \textbf{05}, 135 (2021).

\bibitem{Karndumri:2022rlf}
P.~Karndumri,
``Holographic RG flows and symplectic deformations of N=4 gauged supergravity,''
Phys. Rev. D \textbf{105}, no.8, 086009 (2022).

\bibitem{GW1}
B.P. Abbott et~al., [LIGO Scientific and Virgo Collaborations],
\newblock {GWTC-1: A Gravitational-Wave Transient Catalog of Compact Binary Mergers Observed by LIGO and Virgo during the First and Second Observing Runs},
\newblock {\em Phys. Rev. X} 9(3): 031040, (2019).	




\bibitem{EHT1}
K. Akiyama et~al., [Event Horizon Telescope Collaboration],
\newblock {First M87 Event Horizon Telescope Results. I. The Shadow of the
	Supermassive Black Hole},
\newblock {\em Astrophys. J.} 875(1): L1, (2019).

\bibitem{EHT2}
K. Akiyama et~al., [Event Horizon Telescope Collaboration],
\newblock {First M87 Event Horizon Telescope Results. II. Array and
	Instrumentation},
\newblock {\em Astrophys. J.} 875(1): L2, (2019).

\bibitem{EHT3}
K. Akiyama et~al., [Event Horizon Telescope Collaboration],
\newblock {First M87 Event Horizon Telescope Results. III. Data Processing and
	Calibration},
\newblock {\em Astrophys. J.} 875(1): L3, (2019).

\bibitem{EHT4}
K. Akiyama et~al., [Event Horizon Telescope Collaboration],
\newblock {First M87 Event Horizon Telescope Results. IV. Imaging the Central
	Supermassive Black Hole},
\newblock {\em Astrophys. J.} 875(1): L4, (2019).

\bibitem{EHT5}
K. Akiyama et~al., [Event Horizon Telescope Collaboration],
\newblock {First M87 Event Horizon Telescope Results. V. Physical Origin of the
	Asymmetric Ring},
\newblock {\em Astrophys. J.} 875(1): L5, (2019).

\bibitem{EHT6}
K. Akiyama et~al., [Event Horizon Telescope Collaboration],
\newblock {First M87 Event Horizon Telescope Results. VI. The Shadow and Mass
	of the Central Black Hole},
\newblock {\em Astrophys. J.} 875(1): L6, (2019).

\bibitem{Cunha:2018acu}
P.~V.~P.~Cunha and C.~A.~R.~Herdeiro,
``Shadows and strong gravitational lensing: a brief review,''
Gen. Rel. Grav. \textbf{50}, no.4, 42 (2018).


\bibitem{Lens1}
J.~L.~Synge,
``The Escape of Photons from Gravitationally Intense Stars,''
 Mon. Not. Roy. Astron. Soc. 131(3): 463, (1966).

\bibitem{Lens2}
V.~Bozza,
``Gravitational Lensing by Black Holes'',
 Gen. Rel. Grav. 42: 2269, (2010).


\bibitem{Shape1}
K.~S.~ Virbhadra,
``Relativistic images of Schwarzschild black hole lensing'',
 Phys. Rev. D 79: 083004, (2009).



\bibitem{Shape2}
G.~S.~ Bisnovatyi-Kogan and O.Y. Tsupko,
``Relativistic images of Schwarzschild black hole lensing,''
Plasma Phys. Rep. 41: 562, (2015).






\bibitem{Bar}
J.M. Bardeen, in Black Holes (Proceedings, Ecole d'Et de Physique Thorique: Les Astres Occlus: Les Houches, France, August, 1972)edited by C. DeWitt and B.S. Dewitt.

\bibitem{Shape4}
S.~Chandrasekhar,
The Mathematical Theory of Black Holes (Oxford University Press, New York), (1992).

\bibitem{Shape5}
H.~Falcke, F.~Melia and E.~Agol,
``Viewing the shadow of the black hole at the galactic center,''
 Astrophys. J. Lett. 528: L13, (2000).



\bibitem{Shape6}
V.~Bozza and G.~Scarpetta,
``Strong deflection limit of black hole gravitational lensing with arbitrary source distances,''
Phys. Rev. D 76: 083008, (2007).

\bibitem{Shape7}
C.~Bambi and K.~Freese,
``Apparent shape of super-spinning black holes,''
 Phys. Rev. D 79: 043002, (2009).

\bibitem{Shape8}
P.~G.~Nedkova, V.~K.~Tinchev and S.~S.~ Yazadjiev,
``Shadow of a rotating traversable wormhole,''
 Phys. Rev. D 88(12): 124019, (2013).

\bibitem{spherical1}
R.~Narayan, M.~D.~Johnson and C.~F.~ Gammie,
``The Shadow of a Spherically Accreting Black Hole,''
 Astrophys. J. Lett. 885(2): L33, (2019).



\bibitem{spherical3}
X.~X.~Zeng and H.~Q.~Zhang,
``Influence of quintessence dark energy on the shadow of black hole,''
 Eur. Phys. J. C 80(11): 1058, (2020).



\bibitem{spherical4}
G.~P.~Li and K.~J.~He,
``Observational appearances of a f(R) global monopole black hole illuminated by various accretions,''
 Eur. Phys. J. C 81(11): 1018, (2021).

\bibitem{spherical5}
X.~Qin, S.~B.~Chen and J.~L.~Jing,
``Image of a regular phantom compact object and its luminosity under spherical accretions,''
 Class. Quant. Grav. 38(11): 115008, (2021).

\bibitem{spherical6}
K.~Saurabh and K.~Jusufi,
``Imprints of dark matter on black hole shadows using spherical accretions,''
 Eur. Phys. J. C 81(6): 490, (2021).


\bibitem{spherical7}
X.~X.~Zeng, G.~P.~Li and K.~J.~He,
``The shadows and observational appearance of a noncommutative black hole surrounded by various profiles of accretions,''
 Nucl. Phys. B 974: 115639, (2022).

\bibitem{spherical8}
K.~J.~He, S.~C.~Tan and G.~P.~Li,
``Influence of torsion charge on shadow and observation signature of black hole surrounded by various profiles of accretions,''
 Eur. Phys. J. C 82: 81, (2022).

\bibitem{spherical9}
K.~J.~He, S.~C.~Tan and G.~P.~Li,
``Influence of torsion charge on shadow and observation signature of black hole surrounded by various profiles of accretions,''
Eur. Phys. J. C \textbf{82}, no.1, 81 (2022).

\bibitem{spherical10}
S.~Guo, K.~J.~He, G.~R.~Li and G.~P.~Li,
``The shadow and photon sphere of the charged black hole in Rastall gravity,''
Class. Quant. Grav. \textbf{38}, no.16, 165013 (2021).

\bibitem{spherical11}
X.~X.~Zeng, K.~J.~He and G.~P.~Li,
``Effects of dark matter on shadows and rings of Brane-World black holes illuminated by various accretions,''
Sci. China Phys. Mech. Astron. \textbf{65}, no.9, 290411 (2022).

\bibitem{GMY1}
M. Guo and P-C. Li,
Innermost stable circular orbit and shadow of the 4D Einstein-Gauss-Bonnet black hole,
Eur. Phys. J {\bf C80}, 588 (2020)

\bibitem{GMY2}
X. Wang, P-C. Li, C-Y. Zhang and M. Guo,
Novel shadows from the asymmetric thin-shell wormhole,
Phys. Lett. {\bf B811}, 135930 (2020)

\bibitem{GMY4}
Z. Hu, Z. Zhong, P-C. Li, M. Guo and B. Chen,
QED effect on a black hole shadow,
Phys. Rev. {\bf D103}, 044057 (2021)

\bibitem{SPL}
S.~E.~Gralla, D.~E.~Holz and R.~M.~Wald,
``Black hole shadows, photon rings, and lensing rings,''
 Phys. Rev. D 100(2): 024018, (2019).

\bibitem{thin1}
G.~P.~Li and K.~J.~He,
``Shadows and rings of the Kehagias-Sfetsos black hole surrounded by thin disk accretion,''
JCAP \textbf{06}, 037 (2021).

\bibitem{thin2}
``Q.~Y.~Gan, P.~Wang, H.~W.~Wu and H.~T.~Yang,
Photon ring and observational appearance of a hairy black hole,''
Phys. Rev. D 104(4): 044049, (2021).


\bibitem{Hashimoto:2019jmw}
K.~Hashimoto, S.~Kinoshita and K.~Murata,
``Einstein Rings in Holography,''
Phys. Rev. Lett. \textbf{123}, no.3, 031602 (2019).

\bibitem{Hashimoto:2018okj}
K.~Hashimoto, S.~Kinoshita and K.~Murata,
``Imaging black holes through the AdS/CFT correspondence,''
Phys. Rev. D \textbf{101}, no.6, 066018 (2020).

\bibitem{Liu:2022cev}
Y.~Liu, Q.~Chen, X.~X.~Zeng, H.~Zhang and W.~Zhang,
``Holographic Einstein ring of a charged AdS black hole,''
JHEP \textbf{10}, 189 (2022).

\bibitem{Kaku:2021xqp}
Y.~Kaku, K.~Murata and J.~Tsujimura,
``Observing black holes through superconductors,''
JHEP \textbf{09}, 138 (2021).


\bibitem{Cai:2001dz}
R.~G.~Cai,
``Gauss-Bonnet black holes in AdS spaces,''
Phys. Rev. D \textbf{65}, 084014 (2002).

\bibitem{Glavan:2019inb}
D.~Glavan and C.~Lin,
``Einstein-Gauss-Bonnet Gravity in Four-Dimensional Spacetime,''
Phys. Rev. Lett. \textbf{124}, no.8, 081301 (2020).

\bibitem{Hennigar:2020lsl}
R.~A.~Hennigar, D.~Kubiz\v{n}\'ak, R.~B.~Mann and C.~Pollack,
``On taking the $D\to 4$ limit of Gauss-Bonnet gravity: theory and solutions,''
JHEP \textbf{07}, 027 (2020).

\bibitem{Shu:2020cjw}
F.~W.~Shu,
``Vacua in novel 4D Einstein-Gauss-Bonnet Gravity: pathology and instability?,''
Phys. Lett. B \textbf{811}, 135907 (2020).


\bibitem{Aoki:2020iwm}
K.~Aoki, M.~A.~Gorji and S.~Mukohyama,
``Cosmology and gravitational waves in consistent $D\to 4$ Einstein-Gauss-Bonnet gravity,''
JCAP \textbf{09}, 014 (2020).

\bibitem{spherical2}
X.~X.~ Zeng, H.~Q.~ Zhang and H.~B.~ Zhang,
``Shadows and photon spheres with spherical accretions in the four-dimensional Gauss-Bonnet black hole,''
 Eur. Phys. J. C 80(9): 872, (2020).


\bibitem{Konoplya:2020bxa}
R.~A.~Konoplya and A.~F.~Zinhailo,
``Quasinormal modes, stability and shadows of a black hole in the 4D Einstein\textendash{}Gauss\textendash{}Bonnet gravity,''
Eur. Phys. J. C \textbf{80}, no.11, 1049 (2020)


\end{thebibliography}
\end{document}